\documentclass[pr, twocolumn, preprintnumbers, showpacs, superscriptaddress,longbibliography]{revtex4-1}
\usepackage{amsfonts}
\usepackage{mathrsfs}
\usepackage{graphicx}
\usepackage{amssymb}
\usepackage{color}
\usepackage{amsmath}
\usepackage{graphicx}
\usepackage[
      colorlinks=true,
      linkcolor=blue,
      urlcolor=blue,
      filecolor=black,
      citecolor=blue,
            ]{hyperref}
\usepackage{amsfonts}
\usepackage{amssymb}
\usepackage{epstopdf}
\usepackage{subfigure}
\usepackage{multirow}

\begin{document}
\title{Geodesics connecting end points of time-like interval in asymptotically AdS spacetime}






\author{Jia-Hao He}
\author{Run-Qiu Yang}
\email{aqiu@tju.edu.cn}
\affiliation{Center for Joint Quantum Studies and Department of Physics, School of Science, Tianjin University, Yaguan Road 135, Jinnan District, 300350 Tianjin, P.~R.~China}



\begin{abstract}
It was argued that the holographic entanglement entropy between two timelike separated events in AdS$_3$/CFT$_2$duality needs both timelike and spacelike geodesics. This paper studies the possibility of using smooth spacelike geodesics to connecting two timelike separated events of boundary. We find that in spherically symmetry Schwarzschild AdS black hole, smooth space-like geodesics can connect timelike-separated points by winding around the horizon multiple times. Similar result will also happen in 3-dimensional asymptotically AdS black hole which contains a photon ring in the bulk. Our result suggests that, if there is a photon ring, then the time-like entanglement entropy in the AdS$_3$/CFT$_2$ duality may not have an imaginary part. The result implies that the bulk photon ring may play an important role in understanding analytical behaviors of the entanglement spectrum for AdS$_3$/CFT$_2$ duality.
\end{abstract}
\maketitle

\flushbottom
\section{Introduction}
\label{sec:Intro}
The AdS/CFT correspondence~\cite{Maldacena_1998, Gubser_1998, Witten:1998qj}, also known as holographic principle describes the duality between a ($d+1$)-dimensional gravitational theory in the asymptotically anti-de Sitter spacetime (AdS) and a $d$-dimensional non-gravitational conformal field theory on the boundary. This theory has been used to deal with strongly coupled systems in recent years. Under this correspondence, many related technologies and concepts have been developed, such as holographic entanglement entropy~\cite{Ryu_2006,Nishioka_2009,Ryu:2006bv}, complexity~\cite{Yasuhiro_Sekino_2008,susskind2014entanglement,PhysRevD.90.126007,PhysRevD.93.086006}, etc. These new concepts offer profound insights into the nature of quantum gravity and the solutions of the black hole information paradox~\cite{penington2020entanglement,Almheiri_2020,Almheiri_2021,Calabrese_2004,Srednicki_1993}.

One important application of holographic principles is the computation of correlation functions. However, computing non-perturbative correlation functions in QFT is generally difficult because all loop contributions need to be taken into account. But fortunately, within the framework of AdS/CFT we can use the geodesic approximation~\cite{Balasubramanian_2013,Balasubramanian_2000,Louko_2000,PhysRevD.19.438,Hikida_2022,10.21468/SciPostPhys.15.1.031} method to deal with this problem from the gravity side. According to the holography dictionary, the correlation function in a CFT on the boundary can be expressed as the sum of some propagators in the bulk. Taking the two-point function as an example, we have
\begin{equation}
\label{geodesic_app1}
    \left \langle \mathcal{O}(p)\mathcal{O}(q) \right \rangle=\int \mathcal{D} \mathcal{P} e^{-\frac{\Delta l(\mathcal{P})}{L}}
\end{equation}
where $\{p=\left(\tau_p,\vec{x}_{p} \right),q=\left(\tau_{q},\vec{x}_{q} \right)\}$, are two points on the boundary and $\Delta$ is the conformal dimension of the local operator $\mathcal{O}$. $\mathcal{P}$ is an arbitrary bulk path from $q$ to $p$. $L$ is the AdS radius and $l(\mathcal{P})$ is the length of $\mathcal{P}$. If we let $\Delta \to \infty$, then by saddle-point approximation, (\ref{geodesic_app1}) can be written as
\begin{equation}
\label{geodesic_app2}
    \left \langle \mathcal{O}(p)\mathcal{O}(q) \right \rangle \sim \sum\limits_{\mathrm{geodesics}} e^{-\frac{\Delta l}{L}} \approx e^{-\frac{\Delta l_{min}}{L}}
\end{equation}

The second approximation is because when $\Delta \to \infty$, the $l$ with the smallest real part dominates the result. Through geodesic approximation, we can transform the task of computing correlation function into the calculation of geometric quantities. Nevertheless, this also brings a new question, whether there is such a geodesic between any local operators on the boundary. This issue is also crucial for the study of entanglement entropy, because in the case of $d=3$, the extremal surface in the RT formula correspond to the geodesic in the bulk. In references~\cite{Hawking_Ellis_1973,cmp/1104114862}, they have found that not arbitrary two points in pseudo-Riemannian spacetime can be connected by geodesic. Meanwhile, in those works on time-like entanglement entropy~\cite{Pseudoentropy_timelike_EE,TEE1,LXY_1}, the author found that it seems impossible to find a geodesic that can smoothly connect a pair of time-like points on the time-like boundary.
Similar situations also exist in de Sitter spacetime (dS), where points at space-like separations on its boundary cannot be connected by a single geodesic. Nevertheless, for such points in dS spacetime, we can still calculate their two-point function using the geodesic approximation\cite{Chapman_2023}. This procedure, however, is achieved by employing the wick rotation trick twice, and no actual geodesic connecting these two points has been found. So, does it mean that on the AdS boundary, timelike-separated points cannot be smoothly connected by a geodesic? The answer to this question is not as straightforward as one might think. Indeed, we have found that in the spherically symmetric Schwarzschild-AdS (SAdS) spacetime, there exist bulk geodesics connecting two points on the boundary with time-like separation. The discussions about this problem in the studies of time-like entanglement entropy have only considered simple examples such as pure AdS and BTZ for the case of $d=3$. Moreover, BTZ is locally equivalent to pure AdS. Yet, we find that if there are photon ring/sphere in spacetime, then we can find a geodesic connecting a pair of time-like points on the asymptotically AdS boundary. In certain spacetime under these conditions, timelike-separated points can be connected by either space-like or null geodesics. This suggests that in some 2+1 dimensional asymptotically AdS spacetime, the time-like entanglement entropy may have no imaginary part. In other words, the extremal surface corresponding to a time-like interval on the boundary is smooth and continuous in those spacetime. This phenomenon also suggests that we need further understanding of time-like entanglement entropy.

In order to facilitate description, following the classification of null geodesics in~\cite{Hubeny_2007}, we categorize all geodesics originating from the boundary into the following three types:
Geodesics lying entirely on the boundary ($Type\, \mathcal{A}$), geodesics with only one endpoint on the boundary ($Type\, \mathcal{B}$), and geodesics with both endpoints fixed on the same boundary ($Type\, \mathcal{C}$). Here, it should be emphasized that in certain geometries, spacetime may have multiple boundaries. If the two endpoints of a geodesic lie on different boundaries, we also classify it as a $Type\, \mathcal{B}$ geodesic.
In this work, we mainly focus on the behavior of $Type\, \mathcal{C}$ geodesics, since only this type of geodesic corresponds to extremal surfaces in 2+1-dimensional spacetime. However, what we are truly concerned about is the causal structure possessed by the two endpoints of those geodesics.
Nevertheless, fixing two points on the boundary to directly solve for geodesics can often be quite complicated. Therefore, we can utilize the effective potential method to transform this boundary value problem into a relatively simpler initial value problem.
By leveraging the properties of the effective potential, we can rapidly pick out the $Type\, \mathcal{C}$ geodesics.
To better illustrate this method, as done in~\cite{mustafa2023circular}, we transform this issue into a problem of the motion of free particles starting at the boundary. The effective potential method involves describing the entire spacetime background with an equivalent gravitational potential field. Therefore, in this representation, we only need to find out the conditions under which there exists a potential barrier that can reflect this free particle. Furthermore, through this method, we can also easily calculate the causal structure of both endpoints of $Type\, \mathcal{C}$ geodesics.

The organizational structure of this paper is as follows: In section~\ref{sub:pure_AdS}, we will first use pure AdS as an example to demonstrate some details of the effective potential methods we employed. We will show that in pure AdS, regardless of the space-time's symmetry, the geodesic connecting time-like points on the boundary can not be found. In section~\ref{sec:SW-BH}, we revisit this problem in the SAdS spacetime background. We find that using straightforward periodic identification, geodesic connecting time-like points on the boundary can be found. Due to the complexity of the SAdS solution, in this section, we primarily utilized numerical computation methods. Then in section~\ref{BTZ_modi}, we will conduct a simple verification of our hypothesis under the modified BTZ black hole spacetime background. We find that in spherically symmetric spacetime with photon ring, it is indeed possible for a geodesic to connect two time-like points on the boundary. In 2+1 dimensions, geodesics in the bulk directly correspond to extremal surfaces, which provides us with a deeper understanding of time-like entanglement entropy. In certain geometries, time-like entanglement entropy might be absent of an imaginary component. This implies that we may need further insights into time-like entanglement entropy. In the final section, we will summarize our findings and take a discussion.

\section{Pure AdS}
\label{sub:pure_AdS}
As $d=3$, the length of a bulk geodesic connecting endpoints of an interval on the boundary is equivalent to the area of the extremal surface corresponding to that interval. Although such correspondences do not hold in cases where $d>3$, the conclusions drawn in low-dimensional cases often provide significant insights. Therefore, we first review the calculation of geodesics in pure AdS.
In this section, we use pure AdS as an example to demonstrate some details of the computational methods we employed. We can also see that in pure AdS, there is no possibility of connecting a pair of time-like separated points on the boundary using a geodesic. Usually, the metric of pure $\mathrm{AdS}_{d+1}$ can be written as
\begin{equation}
    \label{eq1.1}
    \mathrm{d}s^2=-f(r) \mathrm{d}t^2+f(r)^{-1} \mathrm{d}r^2 + r^2 \mathrm{d}\boldsymbol{X}^2_{k,d-1}\,
\end{equation}
where $f(r)=k+\frac{r^2}{L^2}$ and $k=0,\,\pm 1$ corresponds to the planar, spherical, and hyperbolic symmetry respectively. $L$ is the AdS radius and we can always make it equal to $1$ by a scaling transformation. $r \in [0, +\infty)$ is the radial coordinate in bulk, and the boundary is located at $r=+\infty$.
$\mathrm{d}\boldsymbol{X}^2_{k,d-1}$ is the metric of the remaining dimensions and
\begin{equation}
    \label{eq1.2}
    \mathrm{d} \boldsymbol{X}^2_{k,d-1}=
    \begin{cases}
        \mathrm{d} \Omega^2_{d-1}=\mathrm{d}\theta^2+\sin^2{\theta} \mathrm{d} \Omega^2_{d-2}\,, \quad k=+1,\\
        \mathrm{d} \mathbf{x}^2_{d-1}=\sum^{d-1}_{i=1} \mathrm{d} x_{i}^2\,, \qquad \qquad \quad k=0,\\
        \mathrm{d} \Theta^2_{d-1}=\mathrm{d}\theta^2+\sinh^2{\theta} \mathrm{d}\Omega^2_{d-2}\,, \quad  k=-1
    \end{cases}
\end{equation}
For the case of $d=3$, we denote $\Omega_1$ and $x_{2}$ as $\phi$. For the sake of simplicity, we can fix $d-2$ degrees of freedom in the spatial directions by a coordinate transformation, leaving only the three free coordinates $t$, $r$, and $\phi$. At the same time, we parameterize the geodesic with an affine parameter $\lambda$, which generally corresponds to the proper length parameter.
Then, by utilizing the Killing vectors, we can define the conserved quantities $E$ and $J$ as follows:
\begin{equation}
    \label{eq1.4}
    E=f(r) \dot t\,\qquad J=r^2 \dot \phi
\end{equation}
The dot means derivation of affine parameter $\lambda$. Let the tangent vector of geodesics be denoted as $u^{a}=\frac{dx^{a}}{d \lambda}=\{u^{t},u^{r},u^{\phi},u^{\theta},\dots\}$. Then the bulk geodesic should satisfy $u_{a} u^{a}=\kappa$, where $\kappa=+1,0,-1$, corresponds to space-like, null and time-like geodesics respectively.
Combining it with the metric \eqref{eq1.1}, we can get
\begin{equation}
    \label{eq1.5}
    -f(r) \dot t^2 + f(r)^{-1} \dot r^2 + r^2 \dot \phi^2 = \kappa
\end{equation}

Then substituting (\ref{eq1.4}) into the above equation, we finally get a differential equation in the radial direction
\begin{equation}
    \label{eq1.6}
    \begin{split}
        \dot r^2 &= \frac{1}{m^2} - V(r)\\
        V(r) &=  \frac{f(r)}{r^2} - \frac{\kappa f(r)}{J^2}
    \end{split}
\end{equation}
Here $V(r)$ is the effective potential of the geodesics. In the above formula, we made a scaling transformation, $\lambda \rightarrow \frac{\lambda}{J}$, and let $\frac{J}{E}=m$.
In \cite{riojas2023photonsphereresponsefunctions}, the parameter $m$ is also referred to as the impact
 parameter. When considering null geodesics, this parameter $m$ alone is sufficient to determine the behavior of the geodesic. However, when considering time-like or space-like geodesics, it is not possible to completely replace both $E$ and $J$ with this parameter $m$. Nevertheless, for consistency, in such cases, we still use $m$ and $J$ as the parameters of the geodesics.

For the two conserved quantities $E$ and $J$ appearing in the above equation, their positive and negative values only represent opposite directions of motion. As we can see, in all expressions involving $E$ and $J$, these two conserved quantities will appear in the form of squares, i.e. $E^2$,  $J^2$, or a combination of them. Then, for simplicity, we can consider only non-negative values. If these two parameters $E$ and $J$ are taken to certain limit values, such as set $J \to 0$, it may also lead to peculiar properties of the effective potential (at least \eqref{eq1.6} is no longer applicable.). Elaborating extensively on this matter here appears overly verbose, so we will place these details in Appendix~\ref{appendix-A}. In the following sections, we will only discuss the most general case where both $E$ and $J$ take a finite positive number.

According to the form of the effective potential in \eqref{eq1.6}, for time-like geodesics, as long as $\lim_{r \to \infty} f(r) = +\infty$, the effective potential $V(r)$ diverges to positive infinity at the boundary. This result leads to the radial geodesic equation \eqref{eq1.6} having no solution. Therefore, in such cases, we are always unable to connect two points on the boundary using time-like geodesics. Pure AdS precisely falls into this scenario. In fact, this conclusion should hold for all asymptotically AdS space-time.
In the subsequent parts of this section, we will proceed to discuss null, and space-like geodesics in order. Additionally, for each type of geodesic, we will separately consider the cases when the spacetime possesses spherical symmetry, planar symmetry, and hyperbolic symmetry, investigating whether corresponding $Type\, \mathcal{C}$ geodesics exist.

\subsection{Null geodesics}
\label{pure-AdS-null}
Firstly, we consider the null geodesics emanating from the boundary. As $r \to \infty$, regardless of the symmetry, we have $V(r) \to 1$. Substituting this into \eqref{eq1.6} and set $\kappa=0$, we obtain $1/m^2-1 = \dot{r}_\infty^2\ge 0$. Thus, irrespective of the symmetry of the spacetime, in order to ensure that at least one point of the null geodesic is fixed on the boundary, we need to set $m \in \left( 0,1\right]$.

For spherical case ($k=1$), its effective potential of null geodesics can be expressed as $V(r)=1+\frac{1}{r^2}$.
In this situation, the effective potential $V(r)$ monotonically decreases, and $V(r) \to \infty$ as $r \to 0$. If the geodesic is able to return to the boundary, then there must be a turning point in bulk, where we have $\dot r=0$. Since the initial ratio of angular momentum to energy $m$ is fixed, the equation obtained by substituting this condition back to (\ref{eq1.6}) can always find a solution. This indicates that in pure AdS we can always find a null geodesic belonging to the $Type\, \mathcal{C}$. On the other hand, since the boundary of a spherically symmetric spacetime is closed, geodesics starting from the boundary will inevitably return to the boundary when there are no singularities or similar structures in bulk. Therefore, this result also aligns with physical expectations.

However, in addition to determining whether a $Type\, \mathcal{C}$ null geodesic exists, we are more concerned about the causality of two points connected by such a geodesic. Then the ratio $|\frac{\Delta t}{\Delta \phi}|$ can be used to determine the causal structure of the intervals located on the boundary. (In the rest of this paper, `$\frac{\Delta t}{\Delta \phi}$' represents `$|\frac{\Delta t}{\Delta \phi}|$' unless stated otherwise.)
Substituting (\ref{eq1.4}) into (\ref{eq1.5}), we can obtain a differential equation only dependent on $r$:
\begin{equation}
    \label{eq1.9a}
    \frac{\mathrm{d}}{\mathrm{d} r}t(r) =\frac{r}{\sqrt{r^{2}-m^{2} f(r)}f(r)}\,,
\end{equation}
and
\begin{equation}
    \label{eq1.9b}
    \frac{\mathrm{d}}{\mathrm{d} r}\phi(r) =\frac{m}{\sqrt{r^{2}-m^{2} f(r)}r}\,.
\end{equation}
We set the coordinates of the starting point of geodesic as $(\infty,t_0,\phi_0)$. With a translation transformation, we can always set the initial condition to $t(\infty)=0$ and $\phi(\infty)=0$.
Then, by integrating \eqref{eq1.9a} and \eqref{eq1.9b}, we can get
\begin{equation}
    \label{eq1.10a}
    t(r) = \arctan\left(\sqrt{\left(-m^{2}+1\right) r^{2}-m^{2}}\right)-\frac{\pi}{2}\,,
\end{equation}
and
\begin{equation}
    \label{eq1.10b}
    \phi(r) = -\arctan\left(\frac{m}{\sqrt{\left(-m^{2}+1\right) r^{2}-m^{2}}}\right)\,.
\end{equation}
The distance from the turning point $r_t$ to the boundary comprises only half of the geodesic; thus, it is imperative to account for a doubling of the result. In this setup, we can get $\Delta t = -2 t(r_{t})$ and $\Delta \phi = -2 \phi(r_{t})$, where $r_t$ satisfies $1/m^2=1+1/r_t^2$.
Then we have $r_{t}=\frac{m}{\sqrt{-m^{2}+1}}$. Further, we can see that $t(r_{t})=-\frac{\pi}{2}$, and $\phi (r_{t})=-\frac{\pi}{2}$. This implies we always have $\frac{\Delta t}{\Delta \phi}=1$. This result indicates that end point of $Type\, \mathcal{C}$ null geodesic will always has a light-like separation. Moreover, it should be noted that $r_t \to \infty$ as $m \to 1$. Since in this limit, $r_t$ will fall onto the boundary, geodesic do not enter the bulk at all. This also means that the geodesic type changes from $Type\, \mathcal{C}$ to $Type\, \mathcal{A}$.

\begin{figure}[h!]
\centering
\subfigure[Spherical]{
\includegraphics[width=0.22\textwidth] {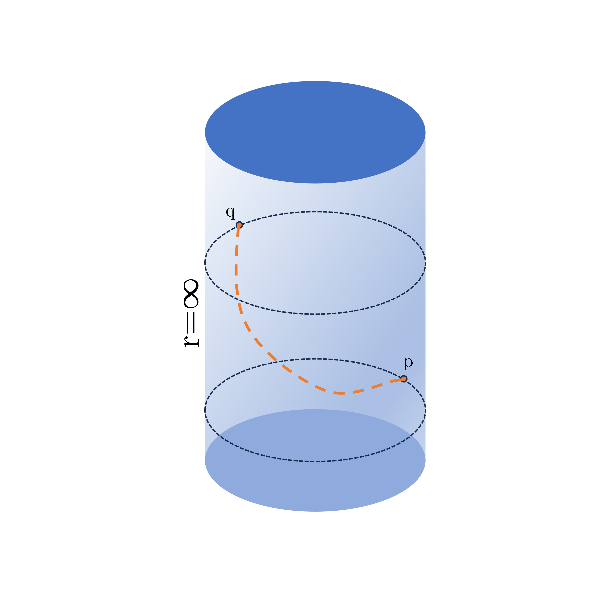}
\label{Fig.shperical_pure_AdS}
}
\subfigure[Planar]{
\includegraphics[width=0.22\textwidth] {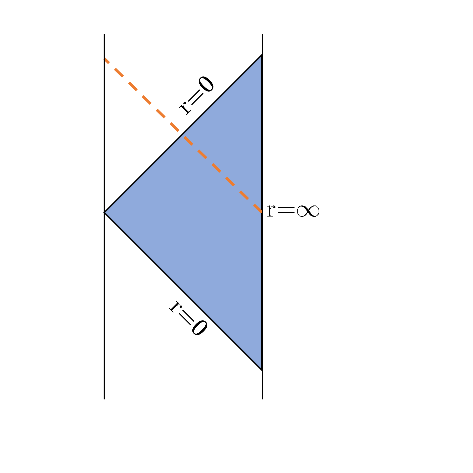}
\label{Fig.planar_pure_AdS}
}
\subfigure[Hyperbolic]{
\includegraphics[width=0.22\textwidth] {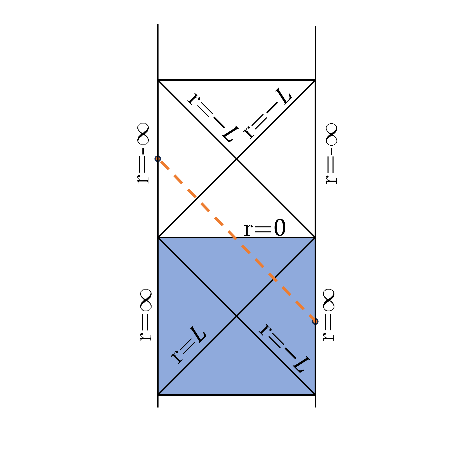}
\label{Fig.hyperbolic_pure_AdS}
}
\caption{A schematic diagram of null geodesics in pure AdS for the different values $k={+1, 0, -1}$. The blue shaded region represents the spacetime that can be covered. The orange dash lines in the figure represent the corresponding null geodesics.}
\label{Fig.vacuum_AdS}
\end{figure}

Then, we generalize the above result to the case of planar($k=0$) and hyperbolic($k=-1$) symmetry.
Here we need to replace the angular momentum with the momentum in the $\phi$ direction, and it does not have periodicity. In addition, $m$ has also been redefined as the ratio of momentum to energy. For pure AdS, although spacetime with three different symmetries are locally equivalent, the ranges covered by their coordinates are different. As shown in figure~\ref{Fig.vacuum_AdS}, in the case of spherical symmetry, the coordinates can cover the maximally extended spacetime, whereas in the cases of planar symmetry and hyperbolic symmetry, only a portion of it can be covered.

Substituting the specific form of $f(r)$ into \eqref{eq1.6}, we can get the corresponding effective potential function : $V(r)=1$ for planar and $V(r)=1-\frac{1}{r^2}$ for hyperbolic. Let us first discuss the case of planar symmetry. Since the effective potential is a constant under planar symmetry, null geodesic departing from the boundary can no longer return back. For all null geodesics with $m < 1$, they will directly reach the position $r = 0$, then extend beyond the range covered by the coordinates under planar symmetry. However, since such geodesics do not correspond to any extremal surface, we are not concerned with these geodesics. Similar to the case of spherical symmetry, when we set $m=1$, we can only find a null geodesic of $Type\, \mathcal{A}$.

For the cases with hyperbolic symmetry we have $\frac{1}{m^2}=1-\frac{1}{r_t^2}$
and we can get $r_{t}=\frac{m}{\sqrt{m^{2}-1}}$. We note that $0 < m \le 1$ and at the same time we require $r_{t}$ to be a real number. This implies that there is actually no such turning point in the bulk when there is hyperbolic symmetry. In addition, there is no singularity at $r=0$, indicating that the entire spacetime can continue to the region where $r<0$. It is also worth noting that in the $r>0$ region, $\dot{r}^2$ varies monotonically, implying that the direction of motion along the $r$-direction for the geodesic remains unchanged. Therefore, the geodesic will directly pass through the position $r=0$, reaching the other boundary at $r=-\infty$. In this situation, the geodesic endpoints fall on two different boundaries, so we can only find geodesic of $Type\, \mathcal{B}$. We have depicted the aforementioned scenario in figure~\ref{Fig.hyperbolic_pure_AdS}. However, similar to the case of planar symmetry, in this scenario, the newly extended spacetime region exceeds the range that can not be covered by the coordinates of a single hyperbolic symmetric pure AdS. Therefore, these geodesics are also not the ones we are concerned with.

\subsection{Space-like geodesics}
\label{pure-AdS-spacelike}
Now let us move on to the case of space-like geodesics, i.e., set $\kappa=1$. We start with the spherical case and then generalize to the planar and hyperbolic cases. According to (\ref{eq1.6}), the effective potential corresponding to the space-like geodesic is
\begin{equation}
    \label{eq.p-s-1}
    V(r)= 1 + \frac{1}{r^2} - \frac{1+r^2}{J^2}
\end{equation}
Compared to the effective potential of null geodesics, there is now an additional constant parameter $J$. Special cases where $J$ or $E$ equals $0$ will be discussed in Appendix \ref{appendix-A}, while in this section, we only consider cases where both parameters are non-zero. We can see that $\lim_{r \rightarrow \infty} V(r)=-\infty$ and $\lim_{r \rightarrow 0} V(r)=\infty$. At the same time, the effective potential is monotonically decreasing in the range $r \in (0,\infty)$. In addition, since $V(r) \in \mathbb{R}$, according to \eqref{eq1.6}, there is $m \in (0,\infty)$. This behavior of $V(r)$ is similar to that of the effective potential for null geodesics in the case of spherical symmetry, which suggest that all space-like geodesics that depart from the boundary are always able to return to the boundary.

We want to know what two points on the boundary can be connected by such a geodesic, and the causal relationship of them. Substitute (\ref{eq1.4}) back to (\ref{eq1.5}) and set $\kappa=1$, followed by directly integrating it, we eventually obtain:
\begin{equation}
    \label{eq.p-s-3}
    \begin{split}
        t \! \left(r \right) &= -\frac{\arctan \! \left\{\frac{\left[m^{2} \left(r^{2}+1\right)-r^{2}+1\right] J^{2}+m^{2} \left(r^{2}+1\right)}{2 \sqrt{\left(r^{2}+1\right) \left(r -J \right) \left(J +r \right) m^{2}+r^{2} J^{2}}\, J}\right\}}{2}+C_1\\
        \phi \! \left(r \right) &= -\frac{\arctan \! \left\{\frac{\left[\left(r^{2}+2\right) J^{2}-r^{2}\right] m^{2}-r^{2} J^{2}}{2 \sqrt{\left[r^{4}+\left(-J^{2}+1\right) r^{2}-J^{2}\right] m^{2}+r^{2} J^{2}}\, J m}\right\}}{2}+C_2
    \end{split}
\end{equation}
Here, we have set the initial condition to $t(\infty)=0$ and $\phi(\infty)=0$. $C_1$ and $C_2$ are the integration constants. Since ultimately we only need to obtain the difference in the variations in the $t$ and $\phi$ directions, the values of these two parameters are arbitrary.

\begin{figure}[h!]
\centering
\includegraphics[width=0.45\textwidth] {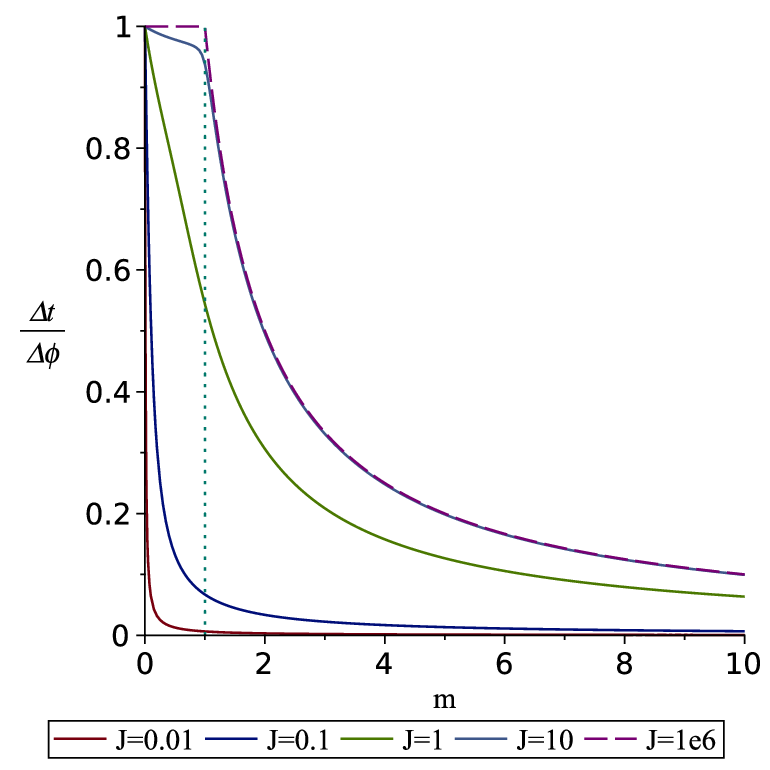}
\caption{The ratio of the distance of the space-like geodesic line in pure AdS in different $J$ for spherical symmetry. We plot curves of $\frac{\Delta t}{\Delta \phi}$ with respect to $m$, with $J=0.01, 0.1, 1, 10, 1\mathrm{e}6$ respectively. When $J \to \infty$, we can see that the curve has a sharp turning point at $m=1$. Within the range of $m<1$, $\frac{\Delta t}{\Delta \phi}$ tends toward $1$, while within the range of $m>1$, it decreases at a rate of $m^{-1}$.}
\label{Fig.3}
\end{figure}

Due to the addition of one more parameter $J$, the solution of the space-like geodesic becomes complicated. When $r \rightarrow \infty$, we have $t(\infty)=-\frac{1}{2}\arctan \! \left[(J^{2} m^{2}-J^{2}+m^{2})/(2 m J)\right]$ and $\phi(\infty)=-\frac{1}{2}\arctan \! \left[(J^{2} m^{2}-J^{2}-m^{2})/(2 m^{2} J)\right]$. Similar to the case of null geodesics, considering the condition $\dot r = 0$, we can then solve \eqref{eq1.6} to find the turning point $r_t$.
Ultimately, we can obtain the interval between the two points connected by the space-like geodesic on the boundary as:
\begin{equation}
    \label{eq.p-s-sph-solutona}
    \Delta t = \frac{\pi}{2}-\arctan\left(\frac{J^{2} m^{2}-J^{2}+m^{2}}{2 m J}\right)\,,
\end{equation}
\begin{equation}
    \label{eq.p-s-sph-solutonb}
    \Delta \phi = \frac{\pi}{2}-\arctan\left(\frac{J^{2} m^{2}-J^{2}-m^{2}}{2 m^{2} J}\right)\,.
\end{equation}
First of all, we notice that there are upper bounds for both $\Delta t$ and $\Delta \phi$, i.e., $\Delta t\le \pi$ and $\Delta \phi \le \pi$. Furthermore, since $\arctan{x}$ is monotonically increasing, we always have $\frac{\Delta t}{\Delta \phi} \le 1 $, which means that the space-like geodesic in pure AdS space-time can only connect space-like points on the boundary. In order to see the effect of the parameter $J$ on the geodesic more clearly, we plot $\frac{\Delta t}{\Delta \phi}$ with respect to $m$ when $J$ take different values, and the results are placed in figure~\ref{Fig.3}.

From figure~\ref{Fig.3} we can see that all curves are monotonically decreasing at $m>0$. And when $J \rightarrow \infty$, $\frac{\Delta t}{\Delta \phi}$ decreases rapidly and tends to 0. This is due to the fact that for each curve we have fixed the value of $J$, and as $m$ increases, the value of $E$ actually decreases, i.e. $m \propto \frac{1}{E}$. Therefore, when $m \rightarrow \infty$, the motion in the $t$ direction will be much smaller than the motion in the $\phi$ direction, so that the interval formed by the two endpoints of the space-like geodesic tends to be pure space-like. Moreover, as $J$ increases, we can see that for $m \in \left( 0,1 \right]$, the curve gradually approaches the solution of the null geodesic, i.e., $\frac{\Delta t}{\Delta \phi}=1$. However, when $m>1$, we can see that the curve will drop rapidly. Especially when $J \rightarrow \infty$, there will exist a sharp inflection point at $m=1$. We plot $J=1\mathrm{e}6$ case to see this more clearly. From (\ref{eq.p-s-sph-soluton}), we can see that when the value of $J$ is very large, the value of the numerator in the $\mathrm{arctan}$ function is dominated by $J$. When $m<1$, the numerator will tend to negative infinity, whereas when $m>1$, the numerator will tend to positive infinity. Therefore, at $m=1$, there will be a jump in the first derivative of $\frac{\Delta t}{\Delta \phi}$. Moreover, we have $\lim\limits_{J \to \infty}\frac{\Delta t}{\Delta \phi}=1$ for $m<1$ and $\lim\limits_{J \to \infty}\frac{\Delta t}{\Delta \phi}=\frac{1}{m}$ for $m>1$.

\begin{figure}[h!]
\centering
\includegraphics[width=0.23\textwidth] {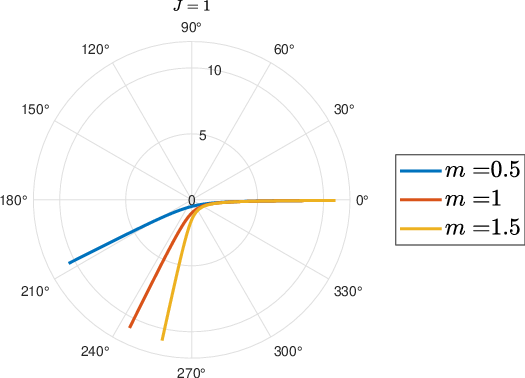}
\includegraphics[width=0.23\textwidth] {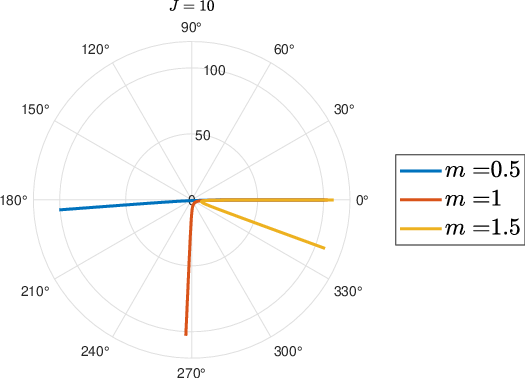}
\includegraphics[width=0.23\textwidth] {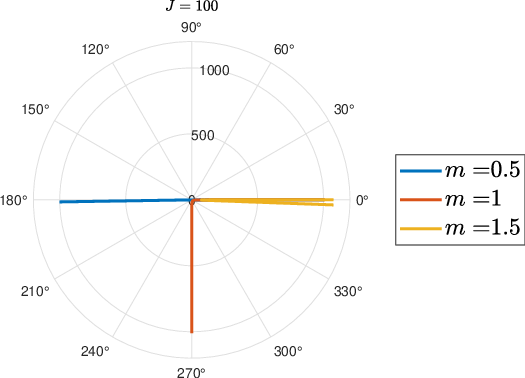}
\includegraphics[width=0.23\textwidth] {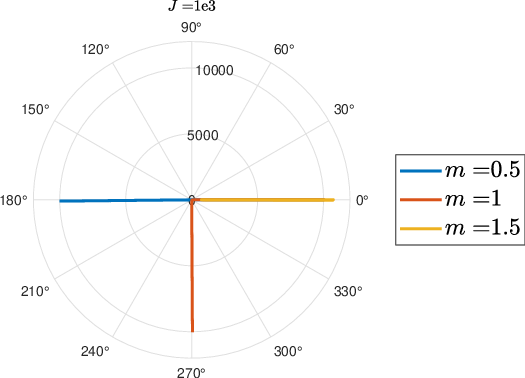}
\caption{The plot of the  space-like geodesic in pure AdS with shperical symmetry. From top to bottom, J={1,10,100,1$\mathrm{e}$3} respectively.}
\label{Fig.pure-space-tr-tphi}
\end{figure}

To further investigate how the geodesics actually changes with the parameter $m$, we plotted the image of the geodesic in the bulk, and the results are in figure~\ref{Fig.pure-space-tr-tphi}.
From the figure, it can be observed that as $J$ increases, the two distinct classes of space-like geodesics, $m<1$ and $m>1$, exhibit different behaviors. Setting one endpoint of those geodesics fixed at coordinates $(t=0, \phi = 0)$ on the boundary, for geodesics with $m<1$, their other endpoint on the boundary gradually approaches $(t=\pi, \phi = \pi)$; whereas for geodesics with $m>1$, their other endpoint approaches $(t=0, \phi = 0)$. According to Eqs.~\eqref{eq.p-s-sph-solutona} and \eqref{eq.p-s-sph-solutonb}, we can draw a conclusion: for any values of $m$ and $J$, it is possible to find a space-like geodesic that connects two points on the boundary with a space-like separation. Furthermore, for any two points on the boundary that are space-like related, there exists a space-like geodesic in the bulk that connects them. This conclusion should hold similarly for spacetime with planar symmetry or hyperbolic symmetry, and we will verify this in the next.

Now, let us check whether the above conclusion still holds true when the spacetime possesses planar symmetry or hyperbolic symmetry.
When space-time has planar symmetry, its corresponding effective potential is $V(r)=1-\frac{r^2}{J^2}$. Using the same condition $\dot r=0$ and combining it with (\ref{eq1.6}), we get $r_{t}=\frac{\sqrt{m^{2}-1}\, J}{m}$. In order to get a real value of the turning point, we can only set $m \ge 1$. In this situation, we have
\begin{equation}
    \label{eq.p-s-pla-solu1}
    \begin{split}
        t \! \left(r \right) &= \frac{-J^{2} m^{2}+m^{2} r^{2}+J^{2}}{\left(m^{2}-1\right) J^{2} \sqrt{-\frac{r^{2} \left(\left(m^{2}-1\right) J^{2}-m^{2} r^{2}\right)}{J^{2}}}} + C_1\\
        \phi \! \left(r \right) &= \frac{-J^{2} m^{2}+m^{2} r^{2}+J^{2}}{\left(m^{2}-1\right) J \sqrt{-\frac{\left(\left(J^{2}-r^{2}\right) m^{2}-J^{2}\right) r^{2}}{m^{2}}}} + C_2
    \end{split}
\end{equation}
Similar to spherical case, $C_1$ and $C_2$ are integral constants and can be omitted. For (\ref{eq.p-s-pla-solu1}), we can take limits at boundary points and turning points respectively and we have
\begin{equation}
    \mathit{\Delta t} = \frac{m}{\left(m^{2}-1\right) J} \qquad \, \Delta \phi = \frac{m^{2}}{\left(m^{2}-1\right) J}
\end{equation}

Then, we have $\frac{\Delta t}{\Delta \phi}=\frac{1}{m}$. We can see that when pure AdS space-time has planar symmetry, the causality of the interval formed by the two ends of the space-like geodesic on the boundary is independent of $J$. Note that in order to find a turning point in the bulk, we have required $m \ge 1$, so we always have $\frac{\Delta t}{\Delta \phi}\le 1$. This also means that in this case, arbitrary two points on the boundary with a space-like separation can be connected by a space-like geodesic.

\begin{figure}[h!]
\centering
\includegraphics[width=0.45\textwidth] {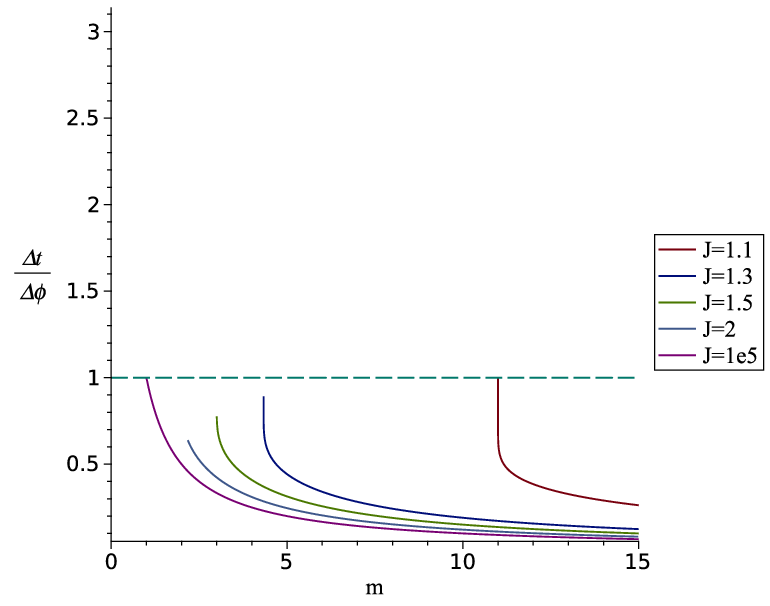}
\caption{The value of $\frac{\Delta t}{\Delta \phi}$ respect to $J$ and $m$ in pure AdS with hyperbolic symmetry. We plot curves of $\frac{\Delta t}{\Delta \phi}$ how varied about $m$, with $J=1.1, 1.3, 1.5, 2, 1\mathrm{e}5$ respectively.}
\label{Fig.p-hy-space-Jm}
\end{figure}

For pure AdS space-time with hyperbolic symmetry, the effective potential of the space-like geodesic depart from the boundary is $V(r)=1-\frac{1}{r^{2}}-\frac{r^{2}-1}{J^{2}}$.
Then similar to the planar case, we can also directly obtain
\begin{equation}
    \label{p-hy-space-inteq1}
    \begin{split}
    &t(r) = -\frac{1}{2}\times\\
    &\mathrm{arctanh}\left\{\frac{2 \sqrt{\left[r^{4}+\left(-J^{2}-1\right) r^{2}+J^{2}\right] m^{2}+J^{2} r^{2}}\, J}{\left[\left(r^{2}-1\right) m^{2}-r^{2}-1\right] J^{2}+\left(-r^{2}+1\right) m^{2}}\right\}\\
    &\phi(r) = -\frac{1}{2}\times\\
    &\mathrm{arctanh}\left\{\frac{2 \sqrt{\left[r^{4}+\left(-J^{2}-1\right) r^{2}+J^{2}\right] m^{2}+J^{2} r^{2}}\, m J}{\left[\left(r^{2}-2\right) m^{2}-r^{2}\right] J^{2}+m^{2} r^{2}}\right\}
    \end{split}
\end{equation}
Expanding the above equation at $r=\infty$, we have $t(\infty)=\mathrm{arctanh}\! \left[(2 J m)/(J^{2} m^{2}-J^{2}-m^{2})\right]$ and $\phi(\infty)=\mathrm{arctanh}\! \left[(2 J \,m^{2})/(J^{2} m^{2}-J^{2}+m^{2})\right]$. Note that at the turning point, the numerator in the arctanh function in (\ref{p-hy-space-inteq1}) is strictly equal to $0$, so the result is just 0. Then we can finally write the analytic expression of $\frac{\Delta t}{\Delta \phi}$ as
\begin{equation}
    \label{p-hy-space-solu}
    \frac{\Delta t}{\Delta \phi}=\left|\frac{\mathrm{arctanh}\left(\frac{2 J m}{J^{2} m^{2}-J^{2}-m^{2}}\right)}{\mathrm{arctanh}\left(\frac{2 J \,m^{2}}{J^{2} m^{2}-J^{2}+m^{2}}\right)}\right|
\end{equation}
The (\ref{p-hy-space-solu}) have been plot in figure~\ref{Fig.p-hy-space-Jm} with different $J$ values. There exists a lower bound $m_{min}=\left| J/(J-1) \right|$ for the parameter $m$ on each curve, and as $m \to m_{min}$, $\frac{\Delta t}{\Delta \phi} \to 1$. In addition, when $J\to 1$, $m\to \infty$. For pure AdS spacetime with hyperbolic symmetry, there exists an event horizon at $r=1$. When $J<1$, the turning point of the corresponding space-like geodesic actually lies inside the event horizon. In this scenario, the geodesic reaches another boundary instead of returning to the original boundary\footnote{Refer to Appendix \ref{appendix-B} for the method to prove this point.}. Such geodesics cannot represent entanglement entropy of interval on the boundary. Therefore, we need to have $J>1$.
For $J>1$,
(\ref{p-hy-space-solu}) is monotonically decreasing and its value is always less than $1$. This result is consistent with the conclusions drawn from spacetime with spherical symmetry: arbitrary two points on the boundary with a space-like separation can be connected by a space-like geodesic.

Let us make a summary. From the calculations in this section, it can be seen that for pure AdS, regardless of its symmetry, only the points with $(\Delta t=\pi, \Delta \phi = \pi)$ on the boundary can be connected by a null geodesic, while arbitrary two spacelike-separated points on the boundary can be connected by a space-like geodesic. Moreover, for two points with a time-like separation in this geometry, there is no smooth geodesic that can connect them. This is consistent with the study of time-like entanglement entropy: in such cases, piece-wise geodesics are required to compute the extremal surface for time-like intervals. However, it is worth questioning whether this situation still holds true for geometries with black holes. Therefore, in the next chapter, we will discuss $Type\, \mathcal{C}$ geodesics in SAdS.

\section{Schwarzschild-AdS}
\label{sec:SW-BH}
In the previous chapter, we calculated various geodesics in pure AdS that can connect two points on the boundary. The results showed that for points with a time-like separation, no smooth geodesic can connect them. However, calculating two-point correlation function with pure time dependence in CFT is not fundamentally difficult. Therefore, to ensure a complete match between the gravitational side and CFT, it is important to explore how to compute correlation functions of two operators with time-like separation on the boundary from the gravitational perspective. Additionally, if we intend to employ the geodesic approximation method, it is essential to have such a geodesic exist. On the other hand, for the extremal surface of time-like interval, a segmented structure has been proposed. If there is no smooth bulk geodesic that can connect two points with time-like separation on the boundary, then in the geodesic approximation, computing the correlation functions of two operators with time-like relation may also require considering such segmented structures. Therefore, attempting to find such a bulk geodesic as described above would be very valuable. For pure AdS, its geometric structure is very simple, and we have only obtained trivial conclusion. However, if a black hole is placed in the spacetime, the situation could be completely different.

In this section, we will take the $\mathrm{SAdS}_4$ spacetime as an example to discuss the causal structure that the endpoints of $Type\, \mathcal{C}$ geodesics may exhibit. The metric of $\mathrm{SAdS}_4$ black hole is given by
\begin{equation}
    \label{eq.swbh-1}
    \begin{split}
    \mathit{\mathrm{d}s}^{2}& = -\left( k+ \frac{r^2}{L^2} -\frac{r_h \left( k+\frac{r_h^2}{L^2} \right)}{r} \right)  \mathit{\mathrm{d}t}^{2}+\\
    &\left( k+ \frac{r^2}{L^2} -\frac{r_h \left( k+\frac{r_h^2}{L^2} \right)}{r} \right)^{-1} \mathit{\mathrm{d}r}^{2} +r^{2} \mathrm{d}\mathbf{X}^{2}_{2}
    \end{split}
\end{equation}
$\mathrm{d}\mathbf{X}^{2}_{2}$ is the same as the definition in \eqref{eq1.2}. In this section we also set $L=1$, and $k=0, 1, -1$ correspond to space-time with planar, spherical, or hyperbolic symmetry, respectively. Similar to the case of pure AdS, the leading term's power of the metric of SAdS spacetime is greater than $1$. Therefore, according to \eqref{eq1.6}, the effective potential $V(r)$ corresponding to time-like geodesics tends to positive infinity at the boundary. This implies that in this spacetime background, its boundary is also inaccessible for time-like geodesics. Hence, we do not require consider time-like geodesics in this scenario as well. In the subsequent parts of this section, we will proceed to discuss null, and space-like geodesics in order.

\subsection{Null geodesic}
\label{SWBH-null}
There has been extensive research on the behavior of radial geodesics in $\mathrm{S(A)dS}$. \cite{Faruk_2024} discussed how radial null geodesics can be used to probe singularity behavior inside black hole horizons or the spacetime beyond cosmological horizons with $d=4,5,6$.
In reference \cite{Kinoshita:2023hgc}, there is some discussion on null geodesics in $d=4$ SAdS spacetime, and in \cite{Hubeny_2007}, null and space-like geodesics in $d=5$ have been studied, where they have calculated the variations of $\Delta t$ and $\Delta \phi$ with respect to $m$. However, because their focus is not related to points on the boundary with time-like separation, they have not emphasized that the $\phi$ coordinate is actually periodic. Due to the spherical symmetry of spacetime, we have $\phi \sim \phi + 2 n \pi$, where $n \in \mathbb{Z}$. Therefore, when considering the causal structure of intervals which endpoint is connected by $Type\, \mathcal{C}$ geodesics, we should describe them using the physically meaningful coordinate $\hat{\phi} \sim (\phi \, \bmod \, 2 \pi)$. Using the $\hat{\phi}$ coordinate is essential for accurately discussing the interval between two points on the boundary, for the actual causal relationships should be determined by $\Delta t$ and $\Delta \hat{\phi}$, which is crucial for the issues we want to study. We find that if we take this into account, it is indeed possible to find geodesic connecting two points on the boundary with time-like separation in SAdS. Therefore, in the following part, we will briefly review the work of previous studies and demonstrate what new result can be obtained by considering the $\hat{\phi}$ coordinate.

Following these previous studies, we can also place the geodesic on the equatorial plane by a coordinate transformation, that is, fix $\theta=\frac{\pi}{2}$. As a result, it is back to almost the same situation as in the previous sections, except that $f(r)$ has a different form. We can write down the corresponding effective potential $V(r)=1+\frac{1}{r^{2}}-\frac{\mathit{r_h}^{3}+\mathit{r_h}}{r^{3}}$.
Obviously, when $r\rightarrow0$, $V(0)=\infty$, and when $r \rightarrow \infty$, $V(\infty)=1$. Then, according to \eqref{eq1.6}, $m \in (0,1]$. For such a effective potential function, we can see that there exist a local maximum point $3 \left( \mathit{r_h}^{3}+ \mathit{r_h}\right) /2$.

\begin{figure}[h!]
\centering
\includegraphics[width=0.45\textwidth] {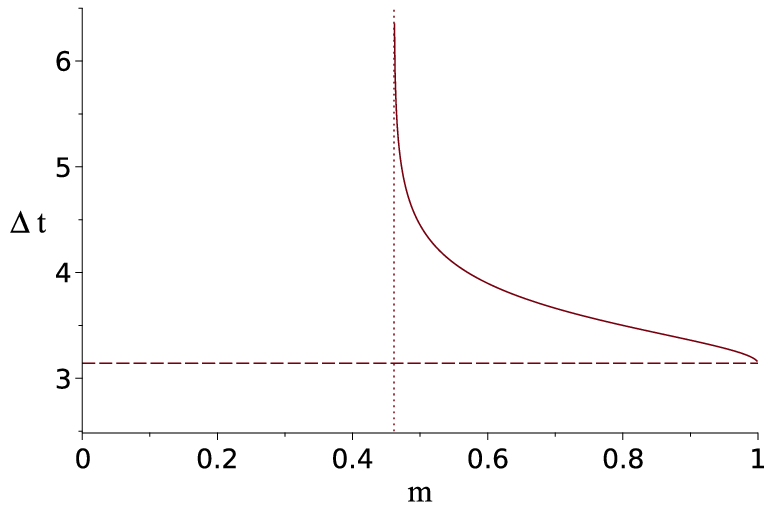}
\includegraphics[width=0.45\textwidth] {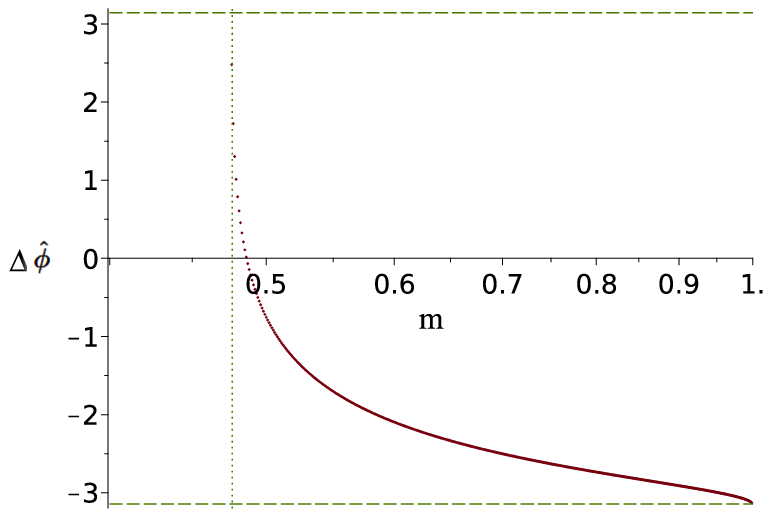}
\caption{The plots of $\Delta t$ and $\Delta \hat{\phi}$ in $\mathrm{SAdS}_4$ with respect to $m$. The horizontal dashed line in the figure is $\Delta t=\pi$ or $\Delta \hat{\phi}=\pm \pi$. The vertical dotted line represents the critical value of $m_c$. When $m$ is less than this threshold, the corresponding geodesics will no longer be able to return to the boundary. We fix $r_h$=0.2, and the critical value here is about $m_c=0.4762$.}
\label{Fig.DT_Dp}
\end{figure}

In the same way, we substitute the new $f(r)$ in \eqref{eq.swbh-1} into equations \eqref{eq1.9a} \eqref{eq1.9b} and use the numerical method to integrate it to obtain the values of $\Delta t$ and $\Delta \hat{\phi}$. The results are shown in figure~\ref{Fig.DT_Dp}. As can be seen in the plot, $\Delta t$ and $\left|\Delta \hat{\phi} \right|$ tend to $\pi$ when $m$ tends to 1. The result finally reduces to the pure AdS situation in this limit. Considering that in such a situation the turning point $r_t$ is rapidly approaching the boundary, and near the boundary the geodesic is not sensitive to the bulk geometry, this phenomenon is reasonable. In addition, if we continue to use the $\phi$ coordinate, there exists a critical point $m_c=\frac{2}{\sqrt{6 \mathit{r_h}^{3}+6 \mathit{r_h}}}$ where both $\Delta t$ and $\Delta \phi$ diverge. On the other hand, considering the turning point $r_t$, when $m \to m_c$, $r \to 3 \left( \mathit{r_h}^{3}+ \mathit{r_h}\right) /2 $. This also implies that the turning point is gradually approaching the location of the photon sphere in the SAdS spacetime. Therefore, theoretically, as $m$ decreases, the geodesic can wind infinitely many times around the photon sphere. Consequently, the $\Delta \phi$ calculated in the $\phi$ coordinate will be much larger than $2 \pi$. Due to the faster growth rate of the \( \phi \) coordinate compared to \( t \), $\frac{\Delta t}{\Delta \phi}\le 1$ is always present.
For those $m<m_c$ cases, the geodesic will directly fall into the event horizon and cannot return.

\begin{figure}[h!]
\centering
\includegraphics[width=0.4\textwidth] {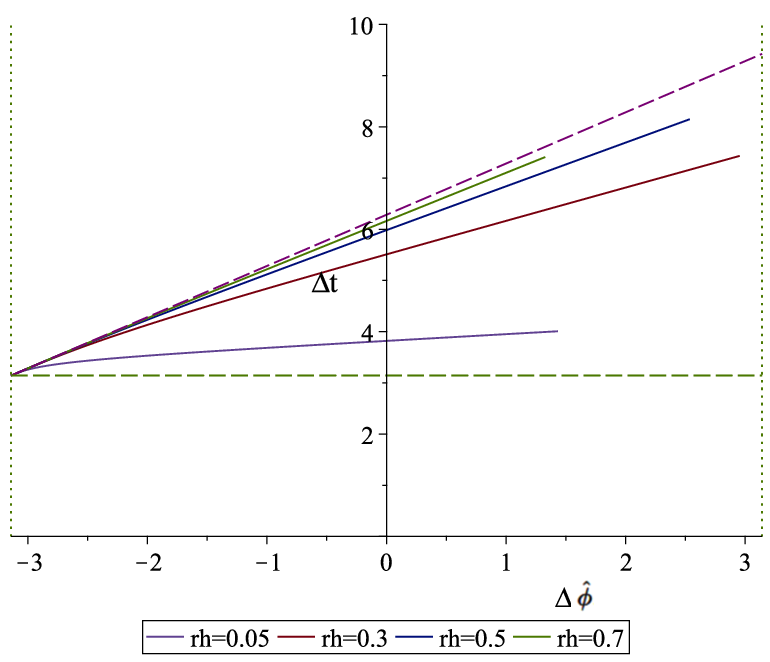}
\caption{The distribution of the two points on the boundary connected by null geodesics originating from the same point. Each solid line in the plot corresponds to a different value of $r_h$. The region swept by each curve represents the points that may be connected by null geodesics at this $r_h$ value. As $m \to 1$, all curves will intersect at the point $\{ \Delta \hat{\phi} = \pm \pi, \Delta t = \pm \pi\}$. The dashed diagonal line in the plot represents the $45^{\circ}$ line. When $r_h \to 1$, those curves gradually approach this line. However, each curve can still extend further, reaching the region above the dashed line in the second period of $\phi$. The dotted lines on the left and right sides represent $\hat{\phi}= \pm \pi$, and these two lines coincide on the cylindrical surface.}
\label{Fig.scanofSWBHnull}
\end{figure}

On the other hand, as long as $m$ is slightly larger than $m_c$, we can always find a turning point near the photon sphere, allowing null geodesics to orbit the photon sphere multiple times before returning to the boundary. In this idea, if we switch to describing it using the physically meaningful coordinate $\hat{\phi}$, it can be found: as long as $\Delta \hat{\phi} < \Delta t$ is satisfied, then the interval it connects just becomes time-like. From figure \ref{Fig.DT_Dp}, it is clear that such a possibility exists. For better observation, in figure \ref{Fig.scanofSWBHnull} we plot the coordinates of two points on the $\Delta t - \Delta \hat{\phi}$ diagram that can potentially be connected by null geodesics. From the figure, we can see that null geodesics can connect points on the boundary that are either time-like or null separated. The region for $\Delta t < \pi$ cannot be covered. It is important to emphasize that figure~\ref{Fig.scanofSWBHnull} actually depicts the computed results for different values of $r_h$. For a fixed $r_h$, it can only show a curve starting from the point $(\pi, \pi)$. Since the length of null geodesics is always zero, the correlation function of the two points connected by such null geodesics according to the geodesic approximation expression always diverges. Based on these results, we can draw a conclusion: For SAdS geometry with spherical symmetry, after fixing the value of $r_h$, only certain special points on the boundary with time-like or light-like separation can be connected by null geodesics.

Now, let us consider the SAdS with planar symmetry, i.e. $k=0$. In this case, the corresponding effective potential is $V(r)=1 -r_h^3/r^3$.  We can see that this effective potential $V(r)$ is a completely monotonically increasing function in the range of $(0,\infty)$. So those null geodesics depart from the boundary cannot return back. This also implies that in this scenario, arbitrary two points on the boundary cannot be connected by a null geodesic.

For hyperbolic symmetrical spacetime, its effective potential is $1-\frac{1}{r^{2}}-\frac{\mathit{r_h}^{3}-\mathit{r_h}}{r^{3}}$. In this case, there is an extreme point $r_c=\frac{3}{2} r_h (1-r_h^2)$ in the effective potential. As $r_h >1$, we have $r_c < 0$, and it is monotonically increasing in the range of $r>0$. When $r\to \infty$, $V(r)\to 1$. Then all geodesic will fall into singularity and no one can return to the boundary.

From the discussion in this subsection, we can conclude: Only in SAdS spacetime with spherical symmetry some special points on the boundary with time-like or null separation can be connected by null geodesics.

\subsection{Space-like geodesic}
\label{SWBH-spacelike}
Then, we move on to the case of space-like geodesic in $\mathrm{SAdS}_4$ spacetime. For the space-like geodesics of the SAdS black hole with $d=5$, there is some relevant discussion in appendix C of \cite{Hubeny_2007}. In their work, they have plotted curves on the $\Delta t - \Delta \phi$ plane under different values of $E$ and $J$, showing points that can potentially be connected by null or space-like geodesics. They observed that the points connectable by null geodesics form a boundary on this plane for those points connectable by space-like geodesics. It is worth emphasizing that while they noted the periodicity of the $\phi$ coordinate, their primary focus was on the observation that points connectable by null geodesics can also be repeatedly connected by multiple space-like geodesics after considering this periodicity. However, by transforming the $\phi$ into a physically meaningful $\hat{\phi} \in [-\pi,\pi]$, we find that this result can even more intriguing than anticipated. With the $\hat{\phi}$ coordinate, we discovered that points with time-like separation can also be connected by space-like geodesics. To clearly demonstrate this, we will briefly present the numerical computation results as before, and show how our new consideration, that is, transforming $\phi$ into $\hat{\phi}$ coordinates, will yield new results.

\begin{figure}[h!]
\centering
\includegraphics[width=0.45\textwidth]{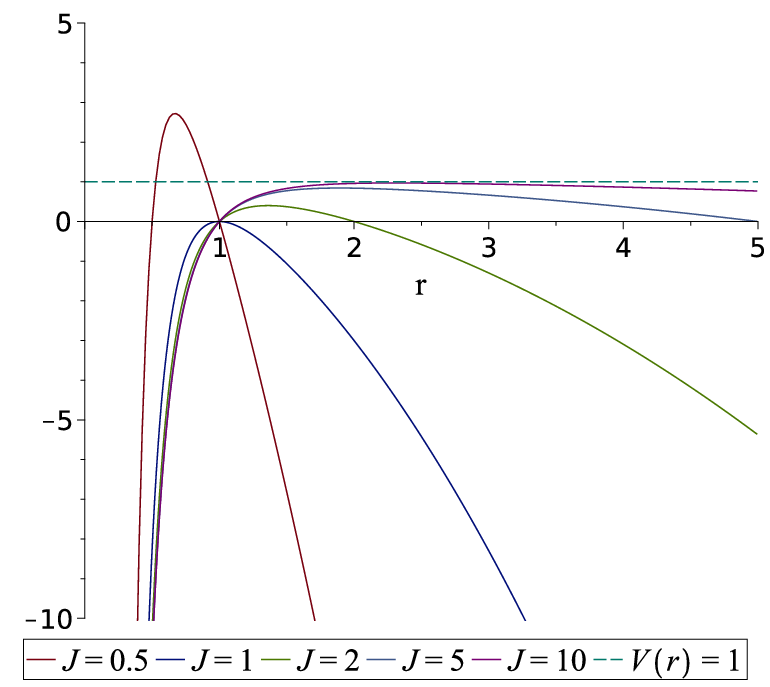}
\includegraphics[width=0.45\textwidth]{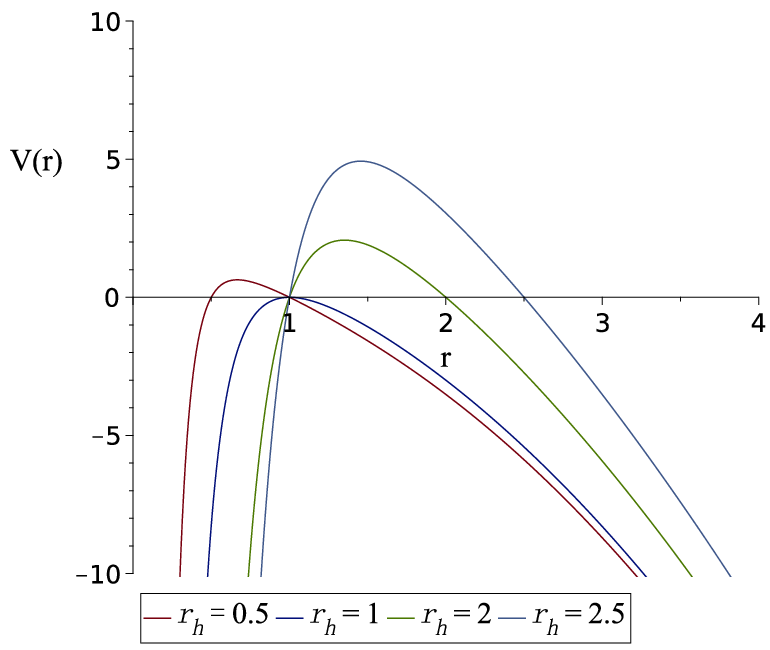}
\caption{The diagram of the effective potential function for space-like geodesic in $\mathrm{SAdS}_4$ space-time. Left: We fix $r_h=1$ and then make $J=\{0.5,1,2,5,10\}$ in order. The horizontal dashed line in the diagram is always equal to 1. Right: $J$ is fixed to 1, then $r_h=\{0.5,1,2,2.5,3\}$}
\label{Fig.9}
\end{figure}

In order to see the properties of the effective potential clearly, we fix the values of $r_h$ and $J$ respectively, and show their influence on the potential function by changing another parameter. The results are plotted in figure~\ref{Fig.9}. As we can see from the diagram, there is only one maximum point. In addition, as $J$ increases, the potential function gradually approaches the horizontal line of $V(r)=1$, and the position of the maximum point is always located between $J$ and $r_h$.

\begin{figure}[h!]
\centering
\includegraphics[width=0.35\textwidth]{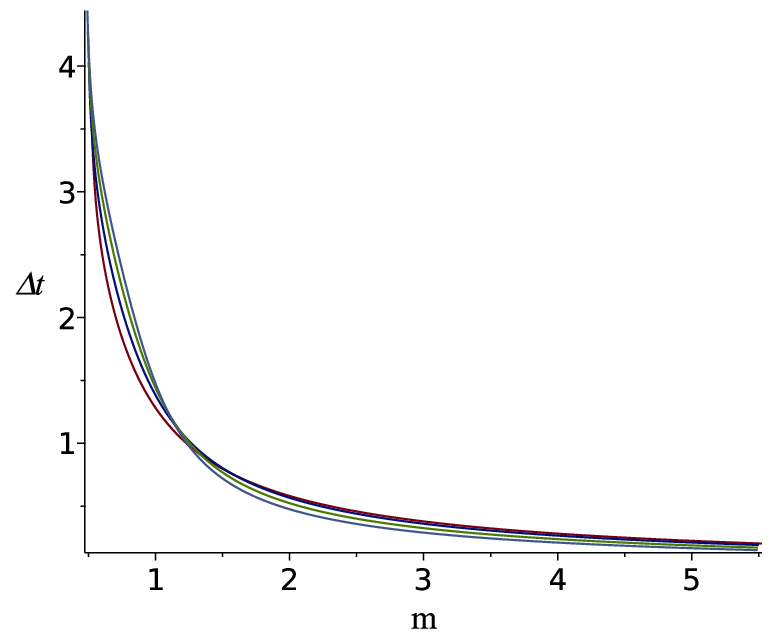}
\includegraphics[width=0.35\textwidth]{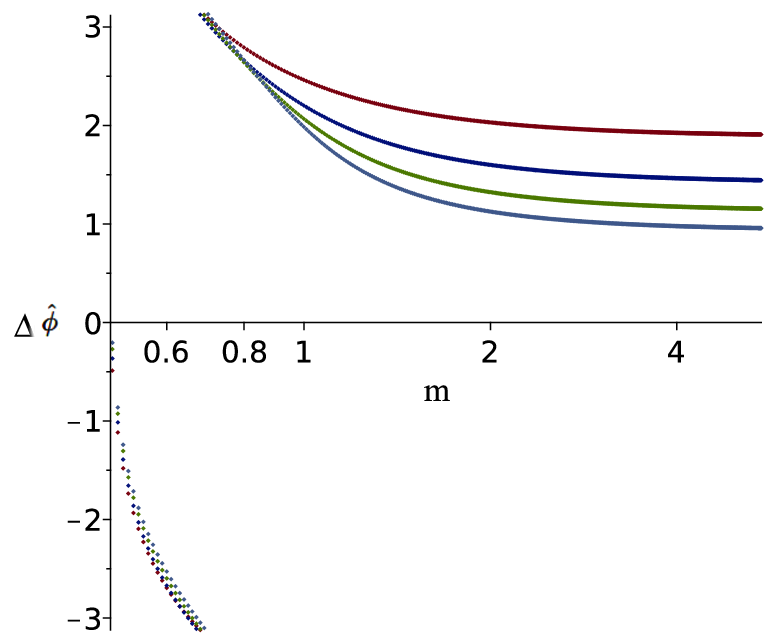}
\caption{The plot of the $\Delta t$ and $\Delta \hat{\phi}$ for the space-like geodesic in $\mathrm{SAdS}_4$ space-time with spherical symmetry. Here we have fix $r_h=0.2$. The value of $J$ decreases from bottom to top, and $J = \{ 0.8,1.2,1.6,2.0 \}$}.
\label{Fig.10}
\end{figure}

It is worth noting that when $J<r_h$, i.e. the turning point is inside the horizon, the space-like geodesics may not return to the boundary where it started, but fall on another boundary. For this phenomenon, \cite{Hubeny_2007} also briefly mentions it, and we will provide a relatively detailed proof of this conclusion in Appendix \ref{appendix-B}.
We are not interested in geodesics that do not return to the same boundary, so in the following parts we only consider the case $J > r_h$.
Similarly to the null case, employing numerical methods, we finally obtain a diagram of $\Delta t$ and $\Delta \hat{\phi}$ about $m$ for the interval connected by the space-like geodesic, the results are plotted in figure~\ref{Fig.10}.

\begin{figure}[h!]
\centering
\includegraphics[width=0.4\textwidth]{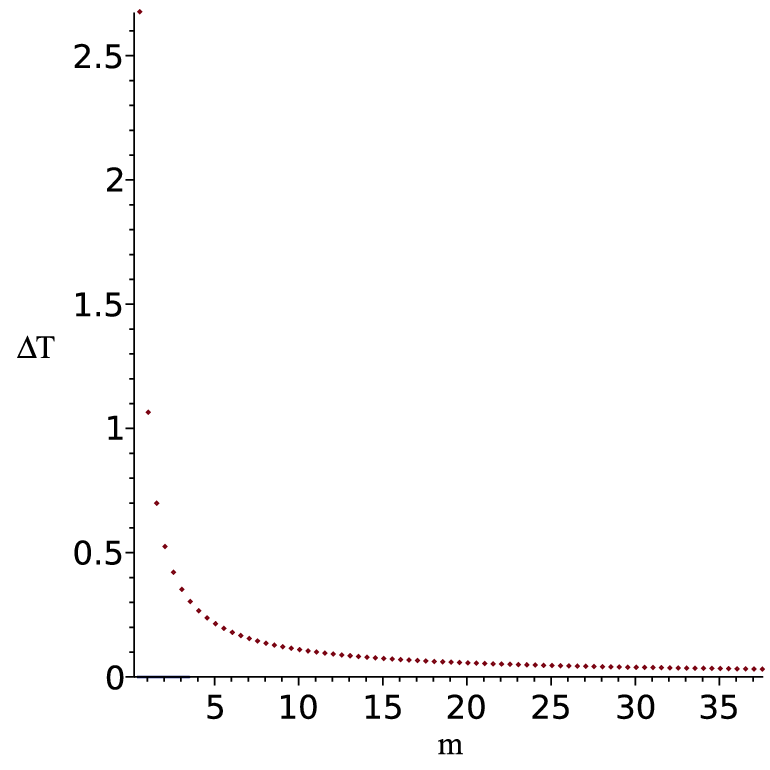}
\includegraphics[width=0.45\textwidth]{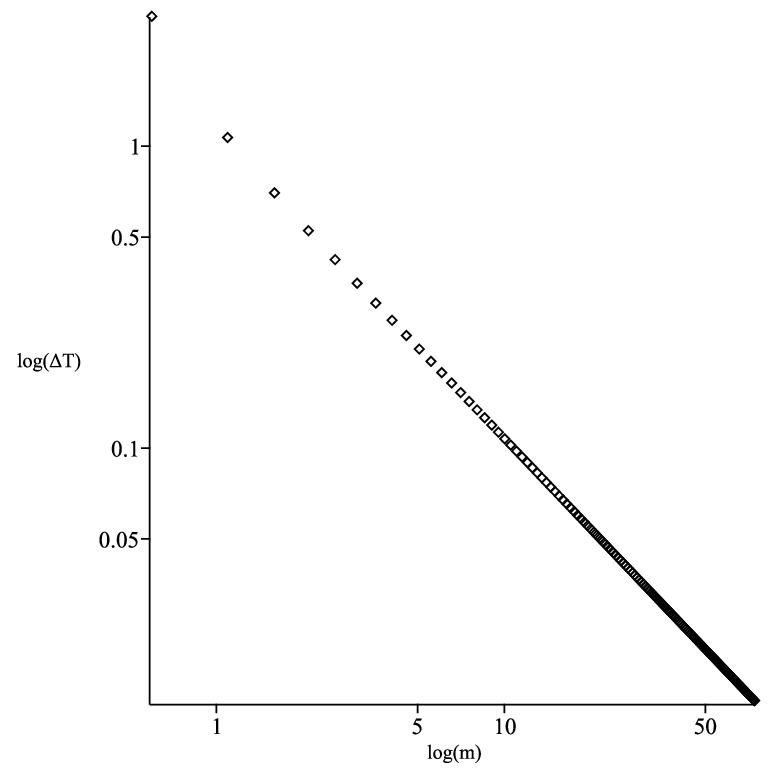}
\caption{The plot of the interval length of a pure time-like interval with appropriate parameters $m$. Each red point in the figure represents a space-like geodesic capable of connecting two purely timelike-separated points on the boundary. We present the same results in a double-logarithmic plot, where it can be observed that they exhibit power law decay. Here, we have fix the $r_h=0.21$.}
\label{Fig.dt-spacelike}
\end{figure}

\begin{figure}[h!]
\centering
\includegraphics[width=0.45\textwidth]{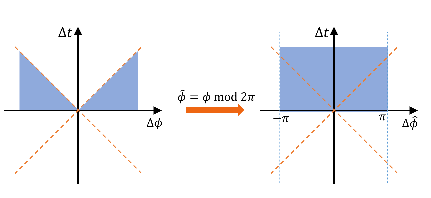}
\caption{The region on the boundary that can be connected by space-like geodesics originating from the origin.}.
\label{Fig.spa-scan-all}
\end{figure}

If we directly compute $\frac{\Delta t}{\Delta \phi}$, we will find that $\frac{\Delta t}{\Delta \phi} < 1$ again. However, after transforming $\phi$ into $\hat{\phi}$, we find that we can directly connect the two points with a timelike-separation on the boundary using space-like geodesics. This brings forth a new question: Does the spherically symmetric SAdS spacetime allow space-like geodesics to connect arbitrary time-like points on its boundary? For the sake of simplicity, let us begin by examining the scenario involving two points with zero spatial separation and only a temporal separation., i.e. set $\Delta \hat{\phi}=0$. After fixing the value of $r_h$, we can indeed find such an space-like geodesic by adjusting the values of $m$ and $J$. The results are presented in figure~\ref{Fig.dt-spacelike}. We set $r_h=0.21$. We find that for any two points on the line that $\hat{\phi}=0$, we can always find a space-like geodesic connecting them. As $m \to \infty$, the $\Delta T$ tends towards $0$. The geodesic obtained in this situation lies on a time slice. This also indicates that two points on the boundary with only a temporal separation can always be connected by a space-like geodesic. Furthermore, can such space-like geodesics cover the entire boundary? We simultaneously fix $\Delta T$ and $\Delta \hat{\phi}$ and perform a grid search in parameter space. Ultimately, we find that at least for any two points in the rectangle region of ${\Delta T \in [0,2 \pi], \Delta \hat{\phi} \in [-\pi,\pi]}$, we can always find such a space-like geodesic solution. To provide a more intuitive explanation of this result, we plot a schematic diagram in figure~\ref{Fig.spa-scan-all}. Although we cannot currently prove that arbitrary two points on the boundary can be connected by a space-like geodesic, based on the results of numerical calculations, we strongly suspect this conclusion to be true.

When using geodesic approximation to compute correlation functions, it is crucial to find a smooth bulk geodesic that connects these two operators. However, in pure AdS, we find that there are no geodesics connecting points with time-like separation, leading us to speculate whether a segmented structure proposed in time-like entanglement entropy might be necessary to address this issue. Nevertheless, in spherically symmetric SAdS spacetime, there are always space-like geodesics connecting two points where $\hat{\phi}=0$. This implies that in certain special geometries, we can directly utilize space-like geodesics to compute time-like correlation functions, without needing to resort to Wick rotation methods to handle this problem. However, for planar symmetric or hyperbolic symmetric SAdS, similar conclusions cannot be drawn.

\begin{figure}[h!]
\centering
\includegraphics[width=0.46\textwidth] {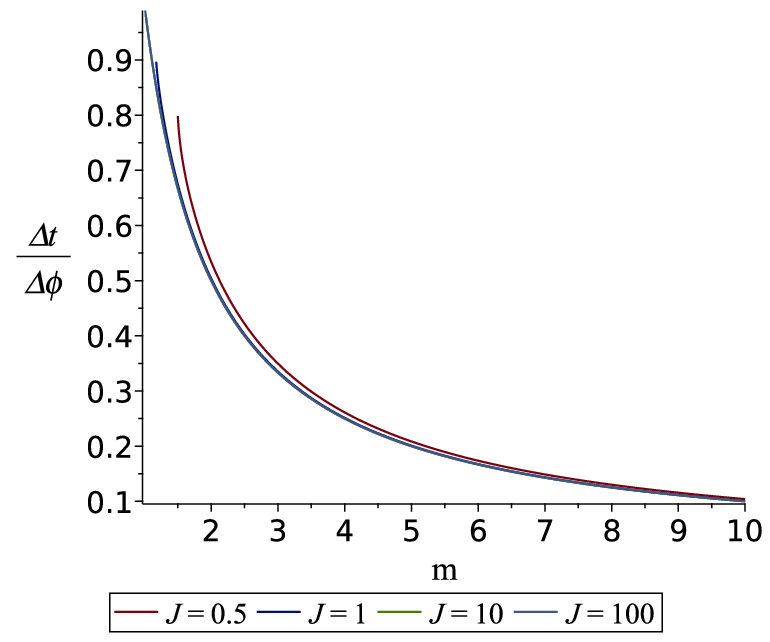}
\includegraphics[width=0.46\textwidth] {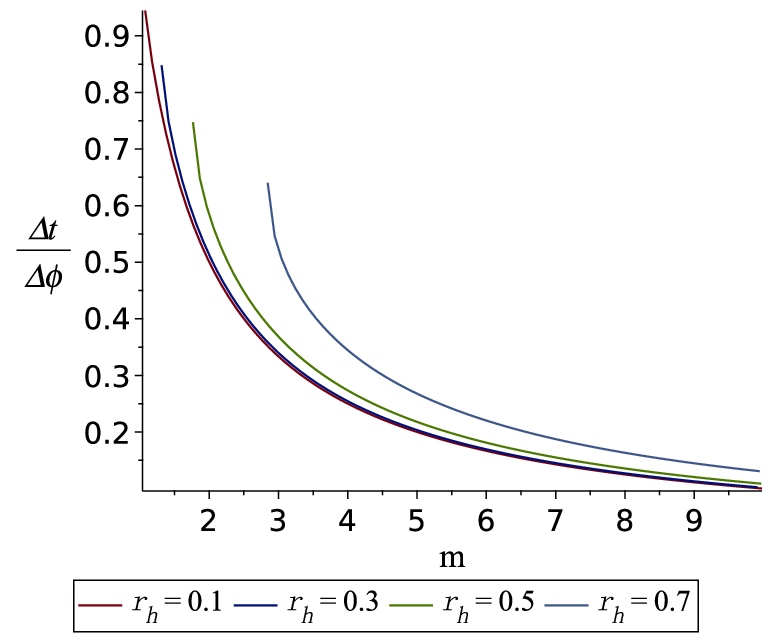}
\caption{The plot of the $\frac{\Delta t}{\Delta \phi}$ for the space-like geodesic in $\mathrm{SAdS}_4$ space-time with planar symmetry. Left: Here we have fix $r_h=0.2$. The value of $J$ increases from top to bottom, and $J = \{ 0.5,1,10,100 \}$. Right: we have fix $J=1$. The value of $r_h$ increases from left to right, and $r_h=\{0.1,0.3,0.5,0.7\} $}
\label{Fig.SW-pla-ratio}
\end{figure}

For SAdS spacetime with planar symmetry, the effective potential corresponding to space-like geodesics is $ V(r)=\left(r^2-\frac{r_h^3}{r} \right) \left( \frac{1}{r^2} - \frac{1}{J^2} \right)$.
Similar to the situation with spherical symmetry, we are also unable to determine the location of the turning point analytically. However, on the other hand, the behavior of the effective potential is almost identical to that of spherical symmetry. At $r=0$ and $r=\infty$, the effective potential tends to negative infinity, while at $r=r_h$ and $r=J$, there exist two zeros and only one maximum point between them. Furthermore, the dependence of the effective potential $V(r)$ on parameters $J$ and $r_h$ is exactly consistent with that shown in figure~\ref{Fig.9}. As $J \to \infty$, $V(\infty) \to  1$, which implies that only $Type\, \mathcal{A}$ geodesics can be found in that limit. In this symmetry, the $\phi$-coordinate no longer exhibits periodicity, thus the ratio $\frac{\Delta t}{\Delta \phi}$ can accurately represent the causal structure of the two points connected by the geodesic. We plot the results of $\frac{\Delta t}{\Delta \phi}$ under planar symmetry in figure~\ref{Fig.SW-pla-ratio}.
Since there is only one maximum value $V(r)_{max}$ in the effective potential, geodesics with $\frac{1}{m^2}>V(r)_{max}$ will directly fall into the singularity without encountering a turning point in the bulk. In figure~\ref{Fig.SW-pla-ratio}, the limit $m \to m_c$ here is $\frac{\Delta t}{\Delta \phi} \to 1$. Therefore, in SAdS spacetime with planar symmetry, space-like geodesics can only connect a pair of points on the boundary that are space-like.

\begin{figure}[h!]
\centering
\includegraphics[width=0.45\textwidth]{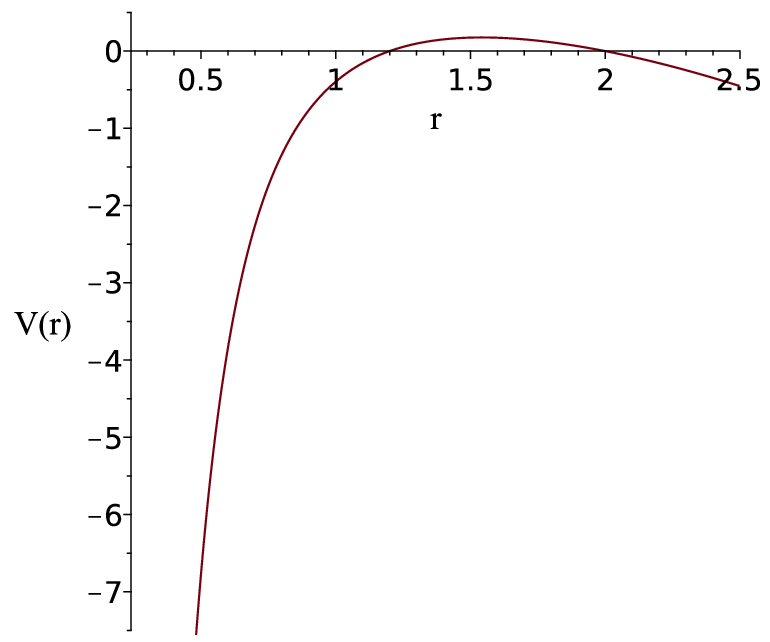}
\includegraphics[width=0.45\textwidth]{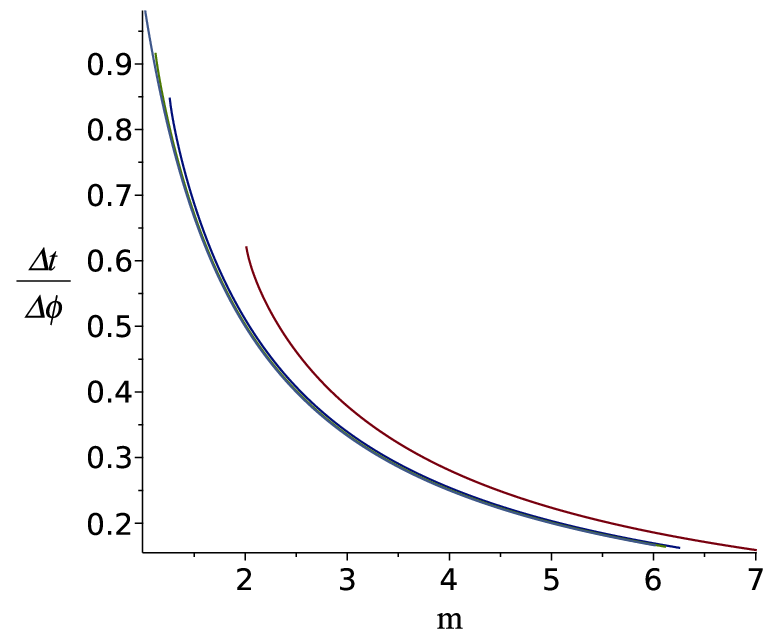}
\caption{Left: The plot of the effective potential $V(r)$ for the space-like geodesic in $\mathrm{SAdS}_4$ space-time with hyperbolic symmetry. We have fix $r_h=1$ and $J=2$. Right: The plot of the $\frac{\Delta t}{\Delta \phi}$ for the space-like geodesic in $\mathrm{SAdS}_4$ space-time with hyperbolic symmetry. The values of $J$ for each curve in the figure decrease from left to right. Here we fix $r_h=1$, and $J=\{ 2, 5, 10, 100 \}$.}
\label{Fig.SW-hy-potential}
\end{figure}

When spacetime possesses hyperbolic symmetry, the behavior of the effective potential is more complicated compared to the previous two cases. Setting $k=-1$, we can express the effective potential as:
\begin{equation}
\label{SW-space-hy-V}
    V(r) = \left( \frac{1}{r^{2}}-\frac{1}{J^{2}} \right) \left[ -1+r^{2}-\frac{\mathit{r_h} \left(\mathit{r_h}^{2}-1\right)}{r} \right]
\end{equation}
The behavior of this effective potential is depicted in the left of figure~\ref{Fig.SW-hy-potential}. Since the behavior of $V(r)$ for $r_h=1$ and $r_h>1$ are similar, we can fix $r_h=1$ without loss of generality. In order to ensure that the turning point $r_t$ is outside the event horizon, we still let $J>r_h$. Due to the $\phi$-coordinate does not have periodicity at this point either, we can directly compute the ratio $\frac{\Delta t}{\Delta \phi}$. We plot the results in the right of figure \ref{Fig.SW-hy-potential}. As $m\to m_c$, $\frac{\Delta t}{\Delta \phi} \to 1$, and all points have $\frac{\Delta t}{\Delta \phi} \le 1$ in the entire range of possible values for $m$. When $m\ge m_c$, all geodesics will directly fall into the singularity. Therefore, we can conclude that in SAdS spacetime with hyperbolic symmetry, only points with space-like separation can be connected by space-like geodesics.

At the end of this section, let us summarize all the discussions above. For spherically symmetric SAdS spacetime, arbitrary two points on its boundary can be connected by a space-like geodesic, while some points with time-like or light-like separation can also be connected by a null geodesic. For cases with planar symmetry or hyperbolic symmetry, only points with space-like separation can be connected by space-like geodesics. Furthermore, these conclusions cannot be directly generalized to all spacetime that contain black holes. For instance, in the non-rotating BTZ black hole, one cannot find geodesics connecting two points on its boundary with time-like separation. However, we notice that spacetime where such geodesics exist have two distinct features: curvature singularity and spherical symmetry. In this geometry, geodesics also involve multiple circuits around a photon sphere to connect points on the boundary with time-like separation. Therefore, we can propose a conjecture: for spacetime that are spherically symmetric and exhibit curvature singularities, or possess unique structures similar to a photon sphere, we can always find points with time-like separation that can be connected by space-like or null geodesics. To further verify this conjecture, we can arbitrarily construct a such spacetime and observe whether it satisfies the conclusion discussed above.

\section{The modified BTZ black hole}
\label{BTZ_modi}
In the previous section, we found that it seems as long as a spherically symmetric spacetime can have a structure allows geodesics to wind around multiple times, then a geodesic that connect any two points on the boundary can be found. We are currently unable to provide a universal proof for this conjecture. However, to verify this hypothesis, we may perform calculations on some spherically symmetric AdS black holes and attempt to find counterexamples. In the field of holographic theory, the BTZ black hole \cite{PhysRevLett.69.1849,Carlip_2005,Ban_ados_1999} has been widely studied due to its low dimension, simplicity in calculations, and other favorable properties. In studies of time-like entanglement entropy, the BTZ black hole is also often used as a typical example for calculations \cite{Pseudoentropy_timelike_EE,TEE1,LXY_1}. However, relevant calculations indicate that on the boundary of the BTZ black hole, two points with time-like separation cannot be connected by a smooth geodesic. Then we need to employ a piece-by-piece method to calculate an appropriate value of time-like entanglement entropy. The main reason for this phenomenon is that the BTZ black hole is locally equivalent to an AdS spacetime, with the only difference being in the topology. In fact, the BTZ black hole can be obtained by identifying appropriately in the AdS spacetime. Furthermore, the BTZ black hole does not have curvature singularities at $r=0$. Since in $d=3$, geodesics in the bulk directly correspond to extremal surfaces, and considering the special characteristics of the BTZ black hole in related calculations, we can consider modifying the BTZ solution to satisfy the conditions outlined in our conjecture. We will then verify the conjecture in the modified geometry.

The metric function for the BTZ black hole (non-rotating) can be expressed as: $f(r)=(r^{2}-r_h^{2})/L^2$. According to the desired CFT, we can consider the boundary of the spacetime with a non-rotating BTZ black hole to be $\mathbb{R}^1 \times \mathbb{R}^1$ or $\mathbb{R}^1 \times \mathbb{S}^1$.
Therefore, we just need make some modifications to the BTZ black hole to introduce genuine singularities:
\begin{equation}
\label{eq.metric_BTZ_modi}
    f(r)=\frac{r^2-r_{\mathrm{m}}^2}{L^2}+\frac{\lambda_0 \ln{\frac{r}{\lambda_0}}}{r}
\end{equation}
The additional term introduced here is an arbitrary handwritten term and does not have any physical significance. The inspiration for this modification comes from the metric of a perfect fluid dark matter black hole (PFDM) \cite{Liang_2023,article_2000_SSPHS,PhysRevD.86.123015}. Here, $\lambda_0$ is a positive real parameter. This parameter significantly affects the range of values for $m$. For simplicity, we will fix it to $\lambda_0=0.1$ later on. Furthermore, since $r_h$ does not represent the position of the event horizon in this case, to avoid ambiguity, we will redefine it as $r_{\mathrm{m}}$. Through such modifications, we not only introduce curvature singularities at $r=0$ in the new spacetime but also ensure that it remains an asymptotically AdS spacetime. Moreover, since \eqref{eq.metric_BTZ_modi} grows faster than $r$, it causes $V(r)$ to diverge towards positive infinity at the boundary when $\kappa = -1$. Therefore, according to \eqref{eq1.6}, we similarly do not need to consider the case of time-like geodesics. Subsequently, we will continue our analysis by examining null and space-like geodesics separately.

\subsection{Null geodesics}
\label{sub:BTZ_modi_null}
\begin{figure}[h!]
\centering
\includegraphics[width=0.45\textwidth]{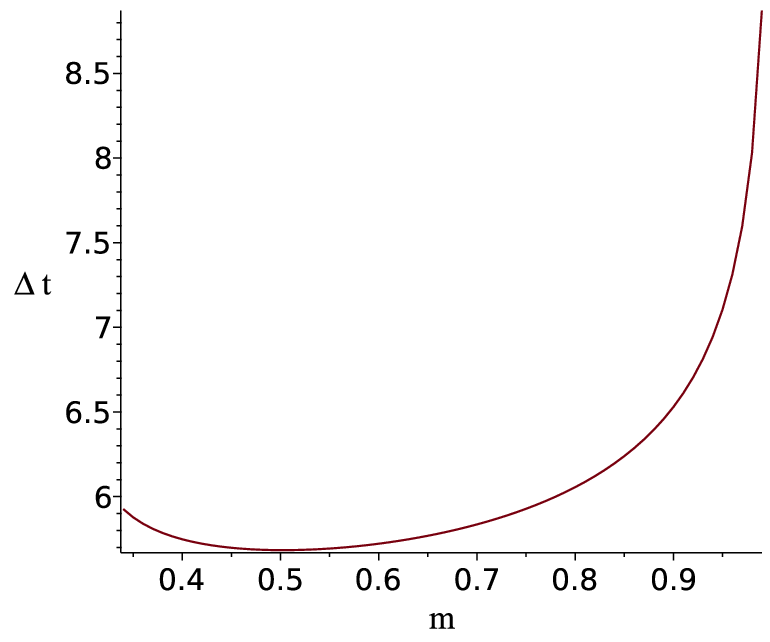}
\includegraphics[width=0.45\textwidth]{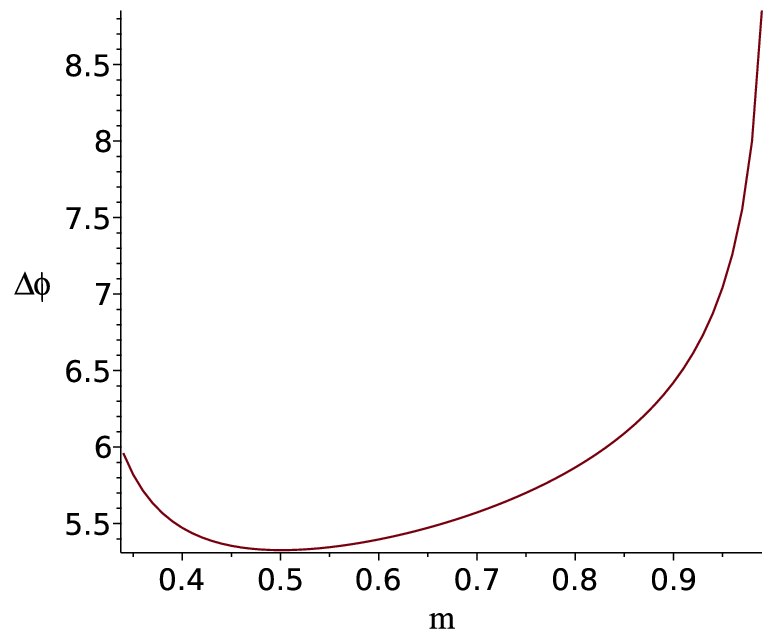}
\caption{The plots of $\Delta t$ and $\Delta \phi$ for null geodesics in the modified BTZ black hole as functions of $m$.}
\label{Fig.BTZ-modi-null-dt-dp}
\end{figure}

The effective potential for null geodesics under this metric is given by:
\begin{equation}
\label{eq.BTZ_moid_potential}
    V(r) = 1 - \frac{r_m^2}{r^2} + \frac{\lambda_0 \ln{\frac{r}{\lambda_0}}}{r^3}
\end{equation}
Manifestly, as $r \to 0$, $V(r) \to -\infty$; as $r \to \infty$, $V(r) \to 1$. Hence, from \eqref{eq1.6}, $m \le 1$. \eqref{eq.BTZ_moid_potential} has two extreme points, one maximum and one minimum. Furthermore, consistent with the previous discussion, we set $r_{m}=0.2$. Utilize the numerical method, we can plot the diagrams of $\Delta t$ and $\Delta \phi$ as functions of $m$, and the results are shown in figure \ref{Fig.BTZ-modi-null-dt-dp}. From the plot, it can be observed that when $m \to 1$, both $\Delta t$ and $\Delta \phi$ exhibit rapid growth behavior. Then, after transforming $\phi$ into $\hat{\phi}$ coordinates, we indeed find some points with time-like separation that can be connected by null geodesics. Moreover, after fixing the parameters $r_{m}$ and $\lambda_0$ in the metric function $f(r)$, there are left with only one free parameter $m$. Therefore, similar to null geodesics in the SAdS black hole, it can only cover a part of points on the boundary.

\subsection{Space-like geodesics}
\label{sub:BTZ_modi_space}
\begin{figure}[h!]
\centering
\includegraphics[width=0.45\textwidth]{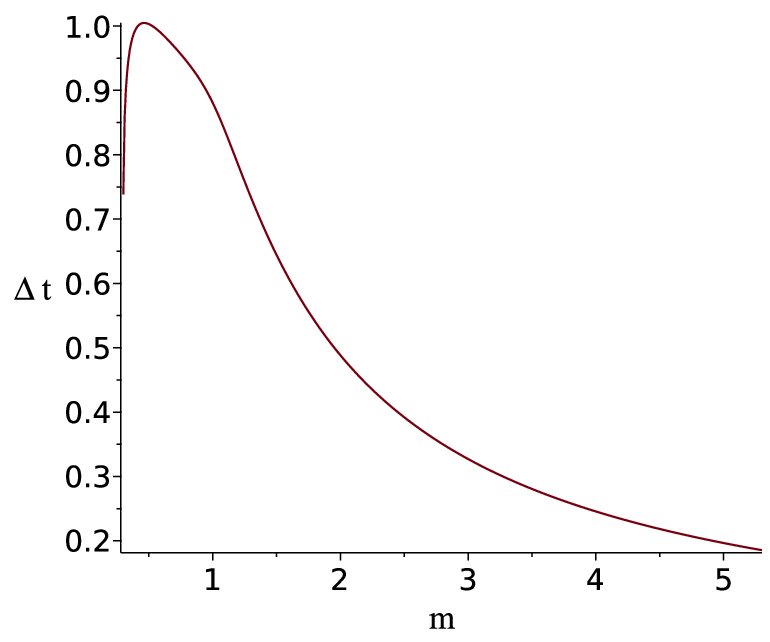}
\includegraphics[width=0.45\textwidth]{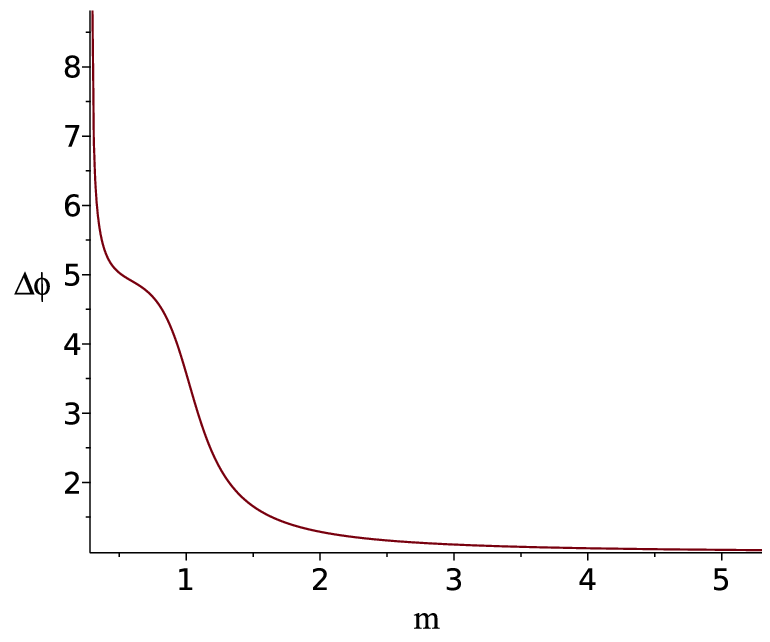}
\caption{The plots of $\Delta t$ and $\Delta \phi$ for space-like geodesics in the modified BTZ black hole as functions of $m$. Here we have set $r_h=0.2$, $\lambda_0=0.1$ and $J=1$.}
\label{Fig.BTZ-modi-space-dt-dp}
\end{figure}

In this subsection, we will investigate the scenario involving space-like geodesics. In the spacetime background described by metric function \eqref{eq.metric_BTZ_modi}, the effective potential $V(r)$ for radial space-like geodesics is given by:
\begin{equation}
\label{eq.BTZ_moid_potential_space}
    V(r)=\left[ r^2 - r_{m}^2 + \frac{\lambda_0 \ln{\frac{r}{\lambda_0}}}{r} \right] \left( \frac{1}{r^2} -\frac{1}{J^2} \right)
\end{equation}
\eqref{eq.BTZ_moid_potential_space} behaves similarly to the effective potential corresponding to space-like geodesics in the spherically symmetric SAdS black hole. It tends to negative infinity as $r\to 0$ and $r\to \infty$, with a maximum at the intermediate region. In this case, we only need to determine the lower bound of $m$ based on this maximum value. To maintain consistency with the discussion about null geodesics, we fix $r_{m}=0.2$. However, in practice, it is imperative to ensure that $r_t \ge r_{m}$. After considering the above factors, the plots of $\Delta t$ and $\Delta \phi$ for different values of $J$ as functions of $m$ are respectively depicted in figure~\ref{Fig.BTZ-modi-space-dt-dp}. Observing the case where $\Delta \phi > 2\pi$ with $\Delta t > 0$ exist, so after transforming $\phi$ into $\hat{\phi}$ coordinates, there indeed have such a situation that satisfying $\frac{\Delta t}{\Delta \hat{\phi}}>1$. In other words, we can find points on the boundary that have time-like separation and can be connected by space-like geodesics.

\section{Conclusion}
\label{Conclu}
In this work, we discussed what kind of points on the boundary of asymptotically AdS spacetime can be connected by smooth geodesics. We conducted analyses on geodesics with different causal structures in three distinct spacetime backgrounds: pure AdS, SAdS, and modified BTZ black holes. Additionally, we considered cases where the spacetime possesses different symmetries. 
In spherically symmetric SAdS spacetime, when $r_h$ is fixed, we find that any two points on the boundary can be connected by space-like geodesics. Particularly, we find that all timelike-separated boundary points can be connected by smooth space-like bulk geodesics in SAdS balck holes. In the case of planar or hyperbolic symmetry for SAdS, however, only points with space-like separation can be connected by space-like geodesics.

Since we have observed that spacetime where two points on the boundary with time-like separation can be connected by geodesics always exhibit photon sphere, we propose a conjecture based on this observation: for spacetime that are spherically symmetric and exhibit a photon sphere/ring, we can always find points with time-like separation that can be connected by space-like or null geodesics. We verified our conjecture in a modified BTZ black hole that satisfies the conditions listed in our conjecture. We found that in this scenario, points on the boundary with time-like separation can indeed be connected by space-like or null geodesics. On the other hand, we found that geodesics connecting points with time-like separation always require the structure similar to a photon sphere/ring, necessitating multiple winding around it. Therefore, we also speculate whether the existence of a unique structure resembling a photon sphere in spacetime is sufficient. A recent paper have shown that in any spherically symmetric spacetime that satisfies asymptotic flatness and possesses an event horizon, a photon sphere always exists \cite{Carballo-Rubio:2024uas}. This proof may perhaps be extended to asymptotically AdS spacetime, suggesting that the condition in our conjecture may be generally satisfied for asymptotically AdS black hole that has spherically symmetry. It is important to emphasize that this conjecture has not yet a universal proof and further verification is necessary.

In previous studies on time-like entanglement entropy \cite{Pseudoentropy_timelike_EE,TEE1,LXY_1}, it has been observed that for a time-like subregion $A$ on the boundary, a single smooth extremal surface cannot be used to connect its boundary $\partial A$. According to our result, in 3-dimensional spacetime that are spherically symmetric and exhibit a photon ring, two time-like separated points on the boundary which can be connected by smooth geodesics seem to always exist. When $d=3$, geodesics are equivalent to extremal surfaces. Therefore, this also implies that in such special geometries, it is possible to have time-like entanglement entropy with only real part and no imaginary part. Moreover, it have been reported that the pseudo entropy of operators with non-Hermitian transition matrices can also be real~\cite{Guo:2022jzs}. Early proposals regarding time-like entanglement entropy were closely linked to pseudo-entropy. Therefore, the discovery of pseudo-entropy in a purely real form aligns with the main conclusions of our work. Thus, it may be possible that in certain special spacetime, calculating time-like entanglement entropy does not require the use of piecewise method. In conclusion, further research and understanding may be needed for time-like entanglement entropy. Particularly, if our conjecture is correct, why does the existence of photon ring play such a special role and what is its dual object in the boundary field thoery when we study timelike holographic entanglement entropy?

\appendix
\section{Extremal behaviors for $J$ and $E$}
\label{appendix-A}

In this section, we will discuss the cases where certain extremal values are taken for $J$ and $E$. We mainly divide it into three cases: $\{E=0, J \not= 0\}$, $\{E\not= 0,J=0\}$, and $\{E=0,J=0\}$.

We need to rewrite (\ref{eq1.6}) as
\begin{equation}
\label{A-1}
    \begin{split}
        \dot r^2 &= E^2 - V(r)\\
        V(r) &=  \frac{J^2 f(r)}{r^2} - \kappa f(r)
    \end{split}
\end{equation}
Moreover, for pure AdS and SAdS spacetime, if there exists an event horizon $r_h$, then $f(r)>0$ for $r>r_h$ always holds. If $r_t < r_h$, the geodesic will not return to the original boundary but will reach another one.

\noindent \textbf{For $\{E=0, J \not= 0\}$ case}

First consider the null geodesic, i.e., $\kappa=0$. Then we have
\begin{equation}
\label{A-2}
    \dot r^2=- \frac{J^2 f(r)}{r^2}<0
\end{equation}
It is clear that there is no solution for geodesics at this situation.

When $\kappa=1$, we have
\begin{equation}
\label{A-3}
    \dot r^2= \frac{(r^2-J^2) f(r)}{r^2}
\end{equation}
In this case there is a turning point $r=J$. In addition, we can also get the differential equations for the $t$ and $\phi$ directions
\begin{equation}
\label{A-4}
    \frac{d}{d r}t \! \left(r \right) = 0 \\, \qquad
    \frac{d}{d r}\phi \! \left(r \right) = \frac{J}{\sqrt{f \! \left(r \right) r^{2}-f \! \left(r \right) J^{2}}\, r}
\end{equation}
From the above equation, it can be easily seen that there is indeed a space-like geodesic connecting two points on the boundary, but these two points must constitute a space-like interval. In this situation, the geodesic lies on a time slice.

When $\kappa=-1$, we have
\begin{equation}
\label{A-5}
    \dot r^2= -\left(f(r)r+ \frac{J^2 f(r)}{r^2}\right)<0
\end{equation}
Again, there is no geodesic solution at this point.
\\ \hspace*{\fill} \\
\noindent \textbf{For $\{E\not= 0,J=0\}$ case}

For the three different types of geodesics, there are:
\begin{equation}
    \dot r^2 =
    \begin{cases}
        E^2\qquad \qquad &\mathrm{For}\quad \kappa=0;\\
        E^2+f(r)\qquad \qquad &\mathrm{For}\quad \kappa=1;\\
        E^2-f(r)\qquad \qquad &\mathrm{For}\quad \kappa=-1.
    \end{cases}
\end{equation}
In both cases of $k = \{0,1\}$, it is clear that there is no turning point. And when $k = -1$, we have $\lim_{r \to \infty} f(r)=\infty$. This implies that the geodesic equation \eqref{A-1} never holds near the boundary. Therefore, there is also no bulk geodesic that can return to the boundary at this time.
\\ \hspace*{\fill} \\
\noindent \textbf{For $\{E=0,J=0\}$ case}

Similarly, we substitute the corresponding parameters into (\ref{A-1}), which can be obtained
\begin{equation}
    \dot r^2 =
    \begin{cases}
        0\qquad \qquad &\mathrm{For}\quad \kappa=0;\\
        f(r)\qquad \qquad &\mathrm{For}\quad \kappa=1;\\
        -f(r)\qquad \qquad &\mathrm{For}\quad \kappa=-1.
    \end{cases}
\end{equation}
Obviously, for null geodesics the turning point is located on the boundary; for space-like geodesics, we can not find a corresponding turning point outside the event horizon. In the case of time-like geodesics, there is also no solution.

\section{Geodesics in Eddington-Finkelstein coordinates}
\label{appendix-B}
In a spacetime with an event horizon $r_h$, consider a space-like geodesic originating from the boundary. Under suitable conditions, this geodesic may have a turning point $r_t$. Although sapce-like geodesics theoretically can return from interior side of the event horizon, if the turning point of the geodesic have $r_t \le r_h$, then in spacetime with maximal extended like SAdS spacetime, it will reach another boundary.
In order to show this, we can handle this problem in in-going Eddington-Finkelstein coordinates. We need to introduce the  tortoise coordinates first, and according to~\cite{Socolovsky2017SchwarzschildBH}, we have
\begin{equation}
\label{tortoise_coordinates}
\begin{split}
    r^{*}(r)=&\int_0^{r} \frac{\mathrm{d}r^{'}}{f(r^{'})}\\
    =&\frac{1}{3 r_h^2+1}\left( r_h \ln{\left|1-\frac{r}{r_h} \right|} -\frac{r_h}{2} \ln{(1+\frac{r(r+r_h)}{r_h^2+1})} \right)\\
    &+\frac{3 r_h^2+2}{\sqrt{3 r_h^2+4}} \mathrm{arctan}(\frac{r\sqrt{3 r_h^2+4}}{2(r_h^2+1)+r r_h})
\end{split}
\end{equation}
Here we have already set the AdS radius $L=1$. And $r^*(r)$ satisfies $r^*(r_h)=-\infty$ and
\begin{equation}
\label{tortoise_limit}
\begin{split}
    r^*(\infty)&=\frac{1}{3 r_h^2 +1} \left( r_h \ln{\sqrt{1+\frac{1}{r_h^2}}} \right.\\
    &+\left.\frac{3 r_h^2+2}{\sqrt{3 r_h^2 +4}} \mathrm{arctan}(\frac{\sqrt{3 r_h^2+4}}{r_h}) \right)
    \end{split}
\end{equation}

With in-going Eddington-Finkelstein coordinates, $v=t+r^*$, we have
\begin{equation}
\label{EF}
    \mathrm{d}s^2=-f(r)\mathrm{d}v^2+ 2\mathrm{d}v \mathrm{d}r+r^2\mathrm{d}\phi^2
\end{equation}

The corresponding geodesic equation under this metric is:
\begin{equation}
\label{gedesic_eq1}
    \begin{split}
        \ddot{v} + \frac{1}{2} f^{'}(r) \dot{v}^2 - r \dot{\phi}^2&=0\\
        \ddot{r}+\frac{1}{2} f(r) f^{'}(r) \dot{v}^2 - rf(r) \dot{\phi}^2-f^{'}(r) \dot{r} \dot{v}&=0\\
        \ddot{\phi}+\frac{2}{r} \dot{\phi} \dot{r}&=0
    \end{split}
\end{equation}
Since only space-like geodesics have the possibility of escaping from the interior of event horizon, we set $\kappa=1$. $E$ and $J$ are the conserved quantities in the $v$ and $\phi$ directions.

The above equation \eqref{gedesic_eq1} can be further simplified into a set of first-order differential equations:
\begin{equation}
    \begin{split}
       \dot{v}=\frac{E+\dot{r}}{f(r)},~\dot{r}=-\frac{\sqrt{E^{2} r^{2}+\kappa  f(r) r^{2}-J^{2} f(r)}}{r},~\dot{\phi}=\frac{J}{r^2}
    \end{split}
\end{equation}
When $r=r_h$, we have $\dot{r}(r_h)=-E$. Then $\dot{v}(r_h)=0$. Recalling that when $\kappa=1$, for the geodesic's turning point to be inside the event horizon, there must have $J<r_t<r_h$. Therefore, within the region of $r_t<r<r_h$, there always exists $E+\dot{r} > 0$, thus $\dot{v} < 0$. This also means that the $v$-coordinate monotonically changes in the interior of the event horizon, causing it to eventually pass through to the other side of the event horizon, rather than returning to its original boundary.

On the other hand, we can also use numerical methods for a double check.
We set the start point at $\{t,r,\phi\}=\{0,0,0\}$. The boundary conditions can be chosen by the effective potential, and we have:
\begin{equation}
\label{BCS-geodesi-eq1}
    \begin{split}
        &v(0)=r^*(\infty),\qquad r(0)=\infty, \qquad \phi(0)=0\\
        &\dot{v}(0)=0,\qquad \dot{r}(0)=-\infty,\qquad \dot{\phi}(0)=0
    \end{split}
\end{equation}
Using numerical methods, it can be found that for a space-like geodesic with a turning point inside the horizon, $v \to -\infty$ after passing the turning point. This also indicates that those geodesics will crosses the event horizon and reach another boundary.

\acknowledgments

This work is supported by the National Natural Science Foundation of China under Grant No. 12375051.


\begin{thebibliography}{10}
\bibitem{Maldacena_1998}
J.M.~Maldacena, \emph{{The Large N Limit of Superconformal Field Theories and Supergravity}}, \href{https://doi.org/10.1023/a:1026654312961}{\emph{Adv. Theor. Math. Phys.} {\bfseries 2} (1998) 231} [\href{https://arxiv.org/abs/hep-th/9711200}{{\ttfamily hep-th/9711200}}].

\bibitem{Gubser_1998}
S.~Gubser, I.~Klebanov and A.~Polyakov, \emph{Gauge theory correlators from non-critical string theory}, \href{https://doi.org/10.1016/s0370-2693(98)00377-3}{\emph{Physics Letters B} {\bfseries 428} (1998) 105-114}.

\bibitem{Witten:1998qj}
E.~Witten, \emph{{Anti-de Sitter space and holography}}, \href{https://doi.org/10.4310/ATMP.1998.v2.n2.a2}{\emph{Adv. Theor. Math. Phys.} {\bfseries 2} (1998) 253} [\href{https://arxiv.org/abs/hep-th/9802150}{{\ttfamily hep-th/9802150}}].

\bibitem{Ryu_2006}
S.~Ryu and T.~Takayanagi, \emph{{Holographic Derivation of Entanglement Entropy from AdS/CFT}}, \href{https://doi.org/10.1103/physrevlett.96.181602}{\emph{Physical Review Letters} {\bfseries 96} (2006) 181602} [\href{https://arxiv.org/abs/hep-th/0603001}{{\ttfamily hep-th/0603001}}].

\bibitem{Nishioka_2009}
T.~Nishioka, S.~Ryu and T.~Takayanagi, \emph{Holographic entanglement entropy: an overview}, \href{https://doi.org/10.1088/1751-8113/42/50/504008}{\emph{Journal of Physics A: Mathematical and Theoretical} {\bfseries 42} (2009) 504008}.

\bibitem{Ryu:2006bv}
S.~Ryu and T.~Takayanagi, \emph{{Holographic derivation of entanglement entropy from AdS/CFT}}, \href{https://doi.org/10.1103/PhysRevLett.96.181602}{\emph{Phys. Rev. Lett.} {\bfseries 96} (2006) 181602} [\href{https://arxiv.org/abs/hep-th/0603001}{{\ttfamily hep-th/0603001}}].

\bibitem{Yasuhiro_Sekino_2008}
Y.~Sekino and L.~Susskind, \emph{Fast scramblers}, \href{https://doi.org/10.1088/1126-6708/2008/10/065}{\emph{Journal of High Energy Physics} {\bfseries 2008} (2008) 065}.

\bibitem{susskind2014entanglement}
L.~Susskind, \emph{Entanglement is not enough},  \href{https://arxiv.org/abs/1411.0690}{{\ttfamily 1411.0690}}.

\bibitem{PhysRevD.90.126007}
D.~Stanford and L.~Susskind, \emph{Complexity and shock wave geometries}, \href{https://doi.org/10.1103/PhysRevD.90.126007}{\emph{Phys. Rev. D} {\bfseries 90} (2014) 126007}.

\bibitem{PhysRevD.93.086006}
A.R.~Brown, D.A.~Roberts, L.~Susskind, B.~Swingle and Y.~Zhao, \emph{Complexity, action, and black holes}, \href{https://doi.org/10.1103/PhysRevD.93.086006}{\emph{Phys. Rev. D} {\bfseries 93} (2016) 086006}.

\bibitem{penington2020entanglement}
G.~Penington, \emph{Entanglement wedge reconstruction and the information paradox},  \href{https://arxiv.org/abs/1905.08255}{{\ttfamily 1905.08255}}.

\bibitem{Almheiri_2020}
A.~Almheiri, R.~Mahajan, J.~Maldacena and Y.~Zhao, \emph{The page curve of hawking radiation from semiclassical geometry}, \href{https://doi.org/10.1007/jhep03(2020)149}{\emph{Journal of High Energy Physics} {\bfseries 2020} (2020) }.

\bibitem{Almheiri_2021}
A.~Almheiri, T.~Hartman, J.~Maldacena, E.~Shaghoulian and A.~Tajdini, \emph{The entropy of hawking radiation}, \href{https://doi.org/10.1103/revmodphys.93.035002}{\emph{Reviews of Modern Physics} {\bfseries 93} (2021) }.

\bibitem{Calabrese_2004}
P.~Calabrese and J.~Cardy, \emph{Entanglement entropy and quantum field theory}, \href{https://doi.org/10.1088/1742-5468/2004/06/p06002}{\emph{Journal of Statistical Mechanics: Theory and Experiment} {\bfseries 2004} (2004) P06002}.

\bibitem{Srednicki_1993}
M.~Srednicki, \emph{Entropy and area}, \href{https://doi.org/10.1103/physrevlett.71.666}{\emph{Physical Review Letters} {\bfseries 71} (1993) 666-669}.

\bibitem{Balasubramanian_2013}
V.~Balasubramanian, A.~Bernamonti, B.~Craps, V.~KerÃ¤nen, E.~Keski-Vakkuri, B.~MÃ¼ller et~al., \emph{Thermalization of the spectral function in strongly coupled two dimensional conformal field theories}, \href{https://doi.org/10.1007/jhep04(2013)069}{\emph{Journal of High Energy Physics} {\bfseries 2013} (2013) }.

\bibitem{Balasubramanian_2000}
V.~Balasubramanian and S.F.~Ross, \emph{Holographic particle detection}, \href{https://doi.org/10.1103/physrevd.61.044007}{\emph{Physical Review D} {\bfseries 61} (2000) }.

\bibitem{Louko_2000}
J.~Louko, D.~Marolf and S.F.~Ross, \emph{Geodesic propagators and black hole holography}, \href{https://doi.org/10.1103/physrevd.62.044041}{\emph{Physical Review D} {\bfseries 62} (2000) }.

\bibitem{PhysRevD.19.438}
L.~Parker, \emph{Path integrals for a particle in curved space}, \href{https://doi.org/10.1103/PhysRevD.19.438}{\emph{Phys. Rev. D} {\bfseries 19} (1979) 438}.

\bibitem{Hikida_2022}
Y.~Hikida, T.~Nishioka, T.~Takayanagi and Y.~Taki, \emph{Cft duals of three-dimensional de sitter gravity}, \href{https://doi.org/10.1007/jhep05(2022)129}{\emph{Journal of High Energy Physics} {\bfseries 2022} (2022) }.

\bibitem{10.21468/SciPostPhys.15.1.031}
L.~Aalsma, M.M.~Faruk, J.P.~van~der Schaar, M.~Visser and J.~de~Witte, \emph{{Late-time correlators and complex geodesics in de Sitter space}}, \href{https://doi.org/10.21468/SciPostPhys.15.1.031}{\emph{SciPost Phys.} {\bfseries 15} (2023) 031}.

\bibitem{Hawking_Ellis_1973}
S.W.~Hawking and G.F.R.~Ellis, \emph{The Large Scale Structure of Space-Time}, Cambridge Monographs on Mathematical Physics, Cambridge University Press (1973).

\bibitem{cmp/1104114862}
B.~Allen and T.~Jacobson, \emph{{Vector two-point functions in maximally symmetric spaces}}, {\emph{Communications in Mathematical Physics} {\bfseries 103} (1986) 669 }.

\bibitem{Pseudoentropy_timelike_EE}
K.~Doi, J.~Harper, A.~Mollabashi, T.~Takayanagi and Y.~Taki, \emph{{Pseudo Entropy in dS/CFT and Time-like Entanglement Entropy}}, \href{https://doi.org/10.1103/physrevlett.130.031601}{\emph{Physical Review Letters} {\bfseries 130} (2023) } [\href{https://arxiv.org/abs/2210.09457}{{\ttfamily 2210.09457}}].

\bibitem{TEE1}
K.~Doi, J.~Harper, A.~Mollabashi, T.~Takayanagi and Y.~Taki, \emph{Timelike entanglement entropy}, \href{https://doi.org/10.1007/jhep05(2023)052}{\emph{Journal of High Energy Physics} {\bfseries 2023} (2023) } [\href{https://arxiv.org/abs/2302.11695}{{\ttfamily 2302.11695}}].

\bibitem{LXY_1}
Z.~Li, Z.-Q.~Xiao and R.-Q.~Yang, \emph{On holographic time-like entanglement entropy}, \href{https://doi.org/10.1007/jhep04(2023)004}{\emph{JHEP} {\bfseries 2023} (2023) } [\href{https://arxiv.org/abs/2211.14883}{{\ttfamily 2211.14883}}].

\bibitem{Chapman_2023}
S.~Chapman, D.A.~Galante, E.~Harris, S.U.~Sheorey and D.~Vegh, \emph{Complex geodesics in de sitter space}, \href{https://doi.org/10.1007/jhep03(2023)006}{\emph{Journal of High Energy Physics} {\bfseries 2023} (2023) }.

\bibitem{Hubeny_2007}
V.E.~Hubeny, H.~Liu and M.~Rangamani, \emph{Bulk-cone singularities \& signatures of horizon formation in ads/cft}, \href{https://doi.org/10.1088/1126-6708/2007/01/009}{\emph{Journal of High Energy Physics} {\bfseries 2007} (2007) 009-009}.

\bibitem{mustafa2023circular}
G.~Mustafa, S.K.~Maurya, A.~Ditta, S.~Ray and F.~Atamurotov, \emph{Circular orbits and accretion disk around ads black holes surrounded by dark fluid with chaplygin-like equation of state},  \href{https://arxiv.org/abs/2311.13839}{{\ttfamily 2311.13839}}.

\bibitem{riojas2023photonsphereresponsefunctions}
M.~Riojas and H.-Y.~Sun, \emph{The photon sphere and response functions in holography},  2023.

\bibitem{Faruk_2024}
M.M.~Faruk, E.~Morvan and J.P.~van~der Schaar, \emph{Static sphere observers and geodesics in schwarzschild-de sitter spacetime}, \href{https://doi.org/10.1088/1475-7516/2024/05/118}{\emph{Journal of Cosmology and Astroparticle Physics} {\bfseries 2024} (2024) 118}.

\bibitem{Kinoshita:2023hgc}
S.~Kinoshita, K.~Murata and D.~Takeda, \emph{{Shooting null geodesics into holographic spacetimes}}, \href{https://doi.org/10.1007/JHEP10(2023)074}{\emph{JHEP} {\bfseries 10} (2023) 074} [\href{https://arxiv.org/abs/2304.01936}{{\ttfamily 2304.01936}}].

\bibitem{PhysRevLett.69.1849}
M.~Banados, C.~Teitelboim and J.~Zanelli, \emph{Black hole in three-dimensional spacetime}, \href{https://doi.org/10.1103/PhysRevLett.69.1849}{\emph{Phys. Rev. Lett.} {\bfseries 69} (1992) 1849}.

\bibitem{Carlip_2005}
S.~Carlip, \emph{Conformal field theory, (2 + 1)-dimensional gravity and the btz black hole}, \href{https://doi.org/10.1088/0264-9381/22/12/r01}{\emph{Classical and Quantum Gravity} {\bfseries 22} (2005) R85-R123}.

\bibitem{Ban_ados_1999}
M.~Banados, \emph{Three-dimensional quantum geometry and black holes}, .

\bibitem{Liang_2023}
X.~Liang, Y.-P.~Hu, C.-H.~Wu and Y.-S.~An, \emph{Thermodynamics and evaporation of perfect fluid dark matter black hole in phantom background}, \href{https://doi.org/10.1140/epjc/s10052-023-12200-8}{\emph{The European Physical Journal C} {\bfseries 83} (2023) }.

\bibitem{article_2000_SSPHS}
Z.~Stuchlik, P.~Slany and S.~HledÃ­k, \emph{Equilibrium configurations of perfect fluid orbiting schwarzschild-de sitter black holes}, {\emph{Astronomy and Astrophysics} {\bfseries 363} (2000) 425}.

\bibitem{PhysRevD.86.123015}
M.-H.~Li and K.-C.~Yang, \emph{Galactic dark matter in the phantom field}, \href{https://doi.org/10.1103/PhysRevD.86.123015}{\emph{Phys. Rev. D} {\bfseries 86} (2012) 123015}.

\bibitem{Carballo-Rubio:2024uas}
R.~Carballo-Rubio and A.~Eichhorn, \emph{{Black hole horizons must be veiled by photon spheres}},  \href{https://arxiv.org/abs/2405.08872}{{\ttfamily 2405.08872}}.

\bibitem{Guo:2022jzs}
W.-z.~Guo, S.~He and Y.-X.~Zhang, \emph{{Constructible reality condition of pseudo entropy via pseudo-Hermiticity}}, \href{https://doi.org/10.1007/JHEP05(2023)021}{\emph{JHEP} {\bfseries 05} (2023) 021} [\href{https://arxiv.org/abs/2209.07308}{{\ttfamily 2209.07308}}].

\bibitem{Socolovsky2017SchwarzschildBH}
M.~Socolovsky, \emph{Schwarzschild black hole in anti-de sitter space}, {\emph{Advances in Applied Clifford Algebras} {\bfseries 28} (2017) } [\href{https://arxiv.org/abs/1711.02744}{{\ttfamily 1711.02744}}].

\end{thebibliography}


%
%
%
%


\end{document}